\documentclass[11pt]{article}

\newcommand{\gaz}{\gamma^0}%

\usepackage{a4}
\usepackage{amssymb,latexsym}
\usepackage[mathscr]{eucal}
\usepackage{citesort}
\usepackage{url}
\usepackage{ntheorem}

\usepackage{mhequ}

\newcommand{\fourrg}{\,{}^{4}g}%
\newcommand{\fourg}{\,{}^{n+1}g}%
\newcommand{\threeg}{h}%

\newcommand{\Nu}{V_u}
\newcommand{\ellek}{l}%

\newcommand{\Jv}[2]{J_{(#1)(#2)}}%
\newcommand{\omv}[1]{\omega_{(#1)}}%
\newcommand{\omvi}{\omv{i}}%
\newcommand{\cv}[1]{c_{(#1)}}%
\newcommand{\jv}[1]{j_{(#1)}}
\newcommand{\jvn}[1]{j_{#1}}%
\newcommand{\Mv}[1]{m_{(#1)}}%
\newcommand{\mv}{\Mv}

\newcommand{\deltak}{\ch}

\newcommand{\Spb}{{\mathfrak S}}

\newcommand{\ch}{\check h}
\newcommand{\ourV}{\mathbb V}

\newcommand{\T}{\Bbb T}
\newcommand{\Mnmo}{{}^{n-1}M}

\newcommand{\ConfInf}{{\mathbb T}^{n-1}}

\newcommand{\hypext}{{\hyp_{\ext}}}

\newcommand{\FS}       
                  {F}

\newcommand{\HS} 
       {H_{\mbox{\scriptsize volume}}}

{\ptc{this should be removed in the oberwolfach version}}%

\newcommand{\zD}{\mathring{D}}%
\newcommand{\zK}{\mathring{K}}%

\newcommand{\ourU}{\mathbb U}%
\newcommand{\eeal}[1]{\label{#1}\end{eqnarray}}
\newcommand{\C}{{\mathbb C}}
\newcommand{\bed}{\begin{deqarr}}
\newcommand{\eed}{\end{deqarr}}
\newcommand{\bedl}[1]{\begin{deqarr}\label{#1}}
\newcommand{\eedl}[2]{\arrlabel{#1}\label{#2}\end{deqarr}}

\newcommand{\mcH}{{\mycal H}}

\newcommand{\mcK}{{\mycal K}}

\newcommand{\bel}[1]{\begin{equation}\label{#1}}
\newcommand{\bea}{\begin{eqnarray}}
\newcommand{\bean}{\begin{eqnarray}\nonumber}
\newcommand{\beal}[1]{\begin{eqnarray}\label{#1}}
\newcommand{\eea}{\end{eqnarray}}

\newcommand{\Eq}[1]{Equation~\eq{#1}}

\newcommand{\qed}{\hfill $\Box$ \medskip}
\newcommand{\proof}{\noindent {\sc Proof:\ }}
\newcommand{\be}{\begin{equation}}
\newcommand{\eeq}{\end{equation}}
\newcommand{\ee}{\end{equation}}
\newcommand{\beqa}{\begin{eqnarray}}
\newcommand{\eeqa}{\end{eqnarray}}
\newcommand{\beqan}{\begin{eqnarray*}}
\newcommand{\eeqan}{\end{eqnarray*}}
\newcommand{\ba}{\begin{array}}
\newcommand{\ea}{\end{array}}

\newcommand{\const}{\mbox{\rm const}} 

\newcommand{\hyp}{\mycal S}

\newcommand{\mcM}{{\mycal M}}
\newcommand{\mcD}{{\mycal D}}

\newcommand{\zf}{\mathring{f}}

\newtheorem{Theorem} {\sc  Theorem\rm} [section]

\newtheorem{Lemma} [Theorem] {\sc  Lemma\rm}
\newtheorem{Proposition} [Theorem] {\sc  Proposition\rm}

\theorembodyfont{\upshape}
\newtheorem{Remark}[Theorem]{\sc  Remark\rm}

\DeclareFontFamily{OT1}{rsfs}{}
\DeclareFontShape{OT1}{rsfs}{m}{n}{ <-7> rsfs5 <7-10> rsfs7 <10-> rsfs10}{}
\DeclareMathAlphabet{\mycal}{OT1}{rsfs}{m}{n}
\def\scri{{\mycal I}}%
\def\Scri{\scri}

{\catcode `\@=11 \global\let\AddToReset=\@addtoreset}
\AddToReset{equation}{section}
\renewcommand{\theequation}{\thesection.\arabic{equation}}

\newcounter{mnotecount}[section]

\renewcommand{\themnotecount}{\thesection.\arabic{mnotecount}}

\newcommand{\mnote}[1]
{\protect{\stepcounter{mnotecount}}$^{\mbox{\footnotesize
$
\bullet$\themnotecount}}$ \marginpar{
\raggedright\tiny\em
$\!\!\!\!\!\!\,\bullet$\themnotecount: #1} }

\newcommand{\warn}[1]
{\protect{\stepcounter{mnotecount}}$^{\mbox{\footnotesize
$
\bullet$\themnotecount}}$ \marginpar{
\raggedright\tiny\em
$\!\!\!\!\!\!\,\bullet$\themnotecount: {\bf Warning:} #1} }

\newcommand{\R}{\mathbb R}

\newcommand{\backn}{{}^{n}b}
\newcommand{\backgn}{{}^{n+1}b}
\newcommand{\eq}[1]{(\ref{#1})}

\newcommand{\ext}{\mathrm{ext}}

\newcommand{\ptM}{\partial\tM}
\newcommand{\tM}{\;\,\,\widetilde{\!\!\!\!\!\mcM}}

\newcommand{\ptc}[1]{\mnote{{\bf ptc:}#1}}

\newcommand{\Ric}{\mbox{\rm Ric}}

\newcommand{\mcL}{{\mycal L}}

\newcommand{\beqar}{\begin{deqarr}}
\newcommand{\eeqar}{\end{deqarr}}

\newcommand{\beaa}{\begin{eqnarray*}}
\newcommand{\eeaa}{\end{eqnarray*}}

\newcommand{\tr}{\mbox{tr}}

\newcommand{\znabla}{\mathring{\nabla}}

\newcommand{\lie}{{\mcL}} 

\newcommand{\zh} {\mathring{h}}

\newcommand{\beq}{\begin{equation}}

\begin{document}
\title{Rigid upper bounds for the angular momentum and centre of mass
of non-singular asymptotically anti-de Sitter space-times}

\author{Piotr T.\ Chru\'sciel\thanks{
E-mail
    \protect\url{Piotr.Chrusciel@lmpt.univ-tours.fr}, URL
    \protect\url{ www.phys.univ-tours.fr/}$\sim$\protect\url{piotr}}
  \\ LMPT,
F\'ed\'eration Denis Poisson, Tours \\
and Albert Einstein Institute, Golm\thanks{Visiting fellow.}
  \\
  \\
 Daniel Maerten\thanks{{
E--mail}: \protect\url{maerten@math.univ-montp2.fr}} \\ Institut
de Math\'ematiques et de Mod\'elisation de Montpellier (I3M)
\\
Universit\'e Montpellier II\\
\\
  Paul Tod\thanks{{ E--mail}: \protect\url{paul.tod@st-johns.oxford.ac.uk}}
\\
Mathematical Institute and St John's College\\ Oxford}


\maketitle


\begin{abstract}
We prove upper bounds on angular momentum and centre of mass in
terms of the Hamiltonian mass and cosmological constant for
non-singular  asymptotically anti-de Sitter initial data sets on
spin manifolds satisfying the dominant energy condition. We work in
space-dimensions larger than or equal to three, and allow a large
class of asymptotic backgrounds, with spherical and non-spherical
conformal infinities; in the latter case, a spin-structure
compatibility condition is imposed. We give classes of non-trivial
examples saturating the inequality. We analyse the borderline case
in space-time dimension four: for spherical cross-sections of Scri,
equality together with completeness occurs only in anti-de Sitter
space-time. On the other hand, in the toroidal case, regular
non-trivial initial data sets saturating the bound exist.
\end{abstract}

\section{Introduction}
\label{introduction}

In recent work~\cite{Maerten} one of us (DM) proved an inequality
satisfied by the global charges for three-dimensional asymptotically
anti-de Sitter initial data sets with spherical conformal infinity.
In this paper we extend that work in several directions by a
consideration of more general initial data sets $(\hyp,g,K)$ on a
spin manifold $\hyp$. The extensions are as follows:
\begin{itemize}
\item
We prove the corresponding inequalities in dimensions $n+1$, $n\ge
3$, with a spin-structure condition for non-spherical Scris
(Theorem~\ref{T1}).
\item
For spherical Scris we obtain optimal inequalities for $n$ equal to
four and five, as well as some natural but non-optimal inequalities
for all $n\geq 3$ (Theorem~\ref{Thgg}; by optimal we mean that
saturation of the inequality is a necessary and sufficient condition
for the existence of space-time Killing spinors at $\hyp$). For
$n=3$, we show that the total momentum cannot be null and give a
sufficient condition slightly stronger than equality for the data to
be anti-de Sitter (Theorem \ref{CTrig2IKS}). Still for $n=3$, when
the associated space-time has a complete Scri with spherical
cross-sections, we prove that equality happens only in anti-de
Sitter space-time (Section~\ref{ssSC}).
\item
For toroidal Scris we obtain optimal inequalities for all $n\ge 3$
(Section~\ref{sSTi}), and we point out the existence of large
families of non-singular (non-vacuum) initial data sets which
saturate the inequality (Sections~\ref{ssTC} and \ref{sssHDT}).

\item We prove that, in dimension $3+1$, with
spherical or toroidal Scri, black hole solutions
saturating the inequality do not exist (Section~\ref{sSBhs}).

\item
We obtain an angular-momentum bound for general conformal boundaries
at infinity with covariantly constant spinors, again under a
spin-structure condition (Section~\ref{SKcb}).
\end{itemize}

\section{Global charges and their positivity}
\label{SGc}

In this work we consider  $n$-dimensional general relativistic data
sets $(\hyp, g, K)$, which are asymptotically anti-de Sitter (adS)
in the following sense: First, we assume that there exists a
Riemannian background metric $b$ which, in the asymptotic region, is
of the form
$$b=dr^2+
\zf(r)\ch\;,
 $$
where $\ch$ is \emph{either} a unit round metric on $S^{n-1}$ (then
with the cosmological constant $\Lambda$ normalised to $-n$, and up
to change of origin in  $r$, $\zf(r)=\sinh^2 r$), \emph{or} $\ch$ is
a Ricci flat metric on an $(n-1)$-dimensional compact manifold
$\Mnmo$ (then, again up to these choices, $\zf(r)=e^{2r}$), where
the space-dimension $n$ is greater than or equal to $3$.
By~\cite{Birmingham}, with those $f$, the initial data $(\hyp,b,0)$
arise from static solutions of vacuum Einstein equations with a
negative cosmological constant\footnote{It might seem natural also
to allow $(\Mnmo,\ch)$ to be a negatively curved Einstein
manifold~\cite{Birmingham}. However, we shall see shortly that such
solutions do not seem to fit into a Witten-type positivity argument,
which is the main concern of this work.}. Note that in the spherical
case, or if $(\Mnmo,\ch)$ is a flat torus $T^{n-1}$, then
$(\hyp,b,0)$ are initial data for anti-de Sitter space-time, or a
quotient thereof. In all cases $(\hyp,b,0)$ provide initial data for
a static Einstein metric.

Next, there is a well-established set of  decay conditions which
guarantee  finite and well defined global charges,
see~\cite{ChHerzlich,ChNagy}, compare~\cite{Wang,Zhang:hpet,HT}.
Following these works, we shall assume that there exist constants
$k\ge 1$, $\alpha>n/2$ and $C>0$ such that for large $r$ we
have\footnote{In \emph{many} of our arguments it is sufficient to
assume the weaker, integral-type, decay conditions
of~\cite[Section~3]{CJL}, but we have not checked whether \emph{all}
the calculations go through under such conditions. }
\bel{falloff} |g-b|_b+|\zD g |_b+\cdots+ |\underbrace{\zD\cdots
\zD}_{k \ \mathrm{factors}} g |_b+|K|_b+\cdots+
|\underbrace{\zD\cdots \zD}_{k -1\ \mathrm{factors}} K |_b\le
Ce^{-\alpha r}\; .
 \ee
Here $|\cdot|_b$ denotes the norm of a tensor field with respect to
the metric $b$, and $\zD$ is the covariant derivative of $b$. These
decay conditions have been chosen because of simplicity of the
analysis involved; it should be recognised that they are
restrictive, and a completely satisfactory treatment should allow
weaker boundary conditions, compare~\cite{CJL} for a related
analysis in the context of a vanishing cosmological constant
$\Lambda=0$.

To define the global charges,  let $X$ be a Killing vector in the asymptotic region of the
associated background space-time. It is well known that each such
$X$ defines a Hamiltonian associated with the flow along
$X$~\cite{HT,ChAIHP,ChNagy,CJL}, as follows: Let $V$ be the normal
component of $X$ with respect to the space-time background metric,
and let $Y$ be the tangential component thereof; when defined along
a spacelike hypersurface, such pairs $(V,Y)$ are called
Killing Initial Data (KIDs). Then the Hamiltonian $H(V,Y)$ corresponding
to $X$ (which we identify with the pair $(V,Y)$) takes the form: \be
\label{mi} H(V,Y):=\lim_{R\to\infty} \frac 1 {16 \pi} \int_{r=R}
\left(\ourU^i(V)+\ourV^i(Y
)\right) dS_i\;,
\ee
where
\begin{eqnarray} \label{eq:3.3} & {}\ourU^i (V):=  2\sqrt{\det
g}\;\left(Vg^{i[k} g^{j]l} \zD_j g_{kl}
+D^{[i}V 
g^{j]k} (g_{jk}-b_{jk}\right) \;,&
\\
& {}\ourV^i (Y):=  2\sqrt{\det
g}\;\left(K^i{_j}-K^k{_k}\delta^i_j\right)Y^j \;.&\label{eq:3.4}
\end{eqnarray}
Here all indices are space indices, running from $1$ to $n$, and
$\zD$ is the Levi-Civita derivative of the space background metric
$b$.

The normalisation constant $1/16\pi$ in \eq{mi} is convenient in
dimension $3+1$ when $\scri$ has spherical cross-section, but
rather arbitrary in higher dimensions, or when non-spherical
cross-sections are considered.

We shall give conditions under which a Witten-type proof of
positivity of global charges applies.   By this we understand an
identity for a spinor $\psi$, relating an appropriate component of
the global charges to an integral over $\hyp$ of a bilinear in
$\psi$; the bilinear is positive given a positivity hypothesis on
the energy-momentum of the initial data set and a differential
equation for $\psi$ at $\hyp$, with suitable asymptotic conditions
for $\psi$ on $\hyp$; positivity of the relevant component of the
global charge follows from a suitable existence theorem for this
differential equation; in the asymptotically-adS setting, we require
$\psi$ to be asymptotic to an \emph{imaginary Killing spinor}, a
notion which we define below. (For more details of the Witten
argument, giving the spinor identity and an existence theorem in
this setting, see e.g. ~\cite{ChHerzlich,Maerten,Wang,CJL}. Examples
in which the Witten argument \emph{does not} apply are given in
\cite{HorowitzMyers,ClarksonMann}.)

Thus the Witten-type proof needs a positivity hypothesis on the
energy-momentum of the initial data set, which will be the
\emph{dominant energy condition} or DEC. Denoting the cosmological
constant by $\Lambda$, we set
$$
\rho:= R-|K|^2+|\tr_g K|^2 -  2 \Lambda\;, \quad J^i =
D_jK^i{_j}-D^i K^j{_j}\;,
$$
where $R$ is the scalar curvature of $g$. The DEC reads then
\bel{posencond}
 \rho \ge |J|_g\;.
 \ee

Next, we  need an \emph{imaginary Killing spinor for the background
metric $b$}; by definition, this is a spinor field $\psi$ in the
asymptotic region $\hypext:=[R_0,\infty)\times \Mnmo$ solving the
set of equations
\bel{iKse}
 \forall \ X\in T\hyp \quad \zD_X \psi = -  i\sqrt{{-\Lambda\over 2n(n-1)}}X \cdot
\psi\;,
 \ee
where $X\cdot$ denotes the Clifford product of $X$, and $\zD$ is the
usual Riemannian spinorial connection associated with the metric
$b$. Such spinor fields are known to exist when $\Mnmo$ is a sphere;
we point out several (well
known~\cite{Leitner,Baum2,Kath2,Bohle,GibbonsRuback}) further
examples, with alternative topologies at infinity, in
Section~\ref{STi} below. The field $\psi$ is a section of a bundle
of spinors\footnote{By a ``spinor field" we mean a section of a
Hermitian bundle associated to the Spin principal bundle over
$\hyp$, equipped with an action of the Clifford algebra of $\hyp$
via anti-Hermitian bundle-morphisms. In what follows we shall freely
make use of ``doubling constructions" such as the one in
\eq{prop2zn} below, and therefore we do \emph{not} impose the
often-implicitly-used condition that the representation of the
Clifford algebra carried by the spinor bundle is irreducible.} which
we will denote by $\Spb'$.

(It is worthwhile pointing out at this stage some more cases when the Witten-type argument cannot be carried through: \emph{manifolds $\hypext$, with
$(\Mnmo,\ch)$ having negative Ricci curvature,  do not admit
imaginary Killing spinors}. This can be seen as follows: first, any
imaginary Killing spinor leads to a Killing vector in $\hypext$. But
it is known, e.g. from the analysis in~\cite[Appendix~A]{ACD2},
that there are no Killing vectors on $\hypext$ in this case. Thus,
no lower bounds on the mass can
 be obtained by Witten-type techniques when, e.g., $\Mnmo$ is a
 two-dimensional higher genus surface.)

For the Witten argument to go through, we need to assume that $\hyp$ admits a spin structure.
Note that the spinor field $\psi$ already singles out a spin
structure on $\hypext$, which is necessarily compatible with the one
of $\hyp$ when $\Mnmo$ is simply connected. However, those spin
structures might be incompatible when $\Mnmo$ is \emph{not} simply
connected. A key, rather restrictive, hypothesis in our work is that
\bel{H1} \mbox{\em the bundle $\Spb'$ over $\hypext$ extends to a
bundle of spinors $\Spb$ over $\hyp$}\;.
 \ee

A short discussion of the hypothesis \eq{H1} is in order. First,
\eq{H1} is satisfied by all product topologies $\hyp=\R\times
\Mnmo$, or $\hyp=[0,\infty)\times \Mnmo$. These examples include the
hyperbolic-cusp solutions \eq{Kottler} below, or  the Kottler black
holes~\cite{Kottler} with toroidal topology at infinity. On the
other hand, \eq{H1} is \emph{not} satisfied by the Horowitz-Myers
solutions~\cite{HorowitzMyers}. Now, in that last example $\hyp$ is
the union of a compact set and of the asymptotic region $\hypext$,
and in such a context we have the following\footnote{We are grateful
to M.~Stern for discussions and references concerning the
compatibility of spin structures.}: If $\Mnmo=\T^2$, the
two-dimensional torus, and $\hyp$ has no boundary (other than the
conformal boundary at infinity), then the trivial spin structure on
$\T^2$, which does admit parallel spinors, \emph{never}
extends~\cite[p.~91]{LawsonMichelsohn} when compactness of the
conformal completion of $\hyp$ is imposed. On the other hand, for
all higher-dimensional toroidal boundaries at infinity
$\Mnmo=\T^{n-1}$, $n\ge 4$, compact boundaryless fillings for the
trivial spin structure of $\T^{n-1}$
exist~\cite[p.~92]{LawsonMichelsohn}. All this leads to a large
class of examples where \eq{H1} holds.

For the analytical arguments to go through, we need further to
assume that $(\hyp,g)$ is \emph{complete}, either without boundary,
or with a
 \emph{compact} boundary satisfying the following: Let $\lambda$ be
 the extrinsic curvature tensor of $\partial \hyp$ (considered as a submanifold of
 $\hyp$, recall that there is no space-time involved at this stage)
 with respect to an inward-pointing unit normal $\nu$, let $h$ be the metric induced on $\partial \hyp$ by $g$.
 The boundary contribution which arises in the Witten argument with a spinor field satisfying the boundary condition of~\cite{GHHP}
 (compare~\cite{Herzlich:mass}) will
have the favorable sign provided that the boundary is either
\emph{weakly future trapped},
\bel{wft} \tr_h \lambda + h^{ab}K_{ab}\le 0\;, \ee
or \emph{weakly past trapped}, which corresponds to changing the
sign in front of the $K$ term in \eq{wft}. An alternative condition
which allows one to conclude is that considered
in~\cite{ChHerzlich,Maerten}. Setting
 $k(\nu)=K_{ia}\nu^idx^a$, where the $x^a$'s are coordinates on $\partial \hyp$, we then assume that
 \bel{traps}
\tr_h\lambda +|k(\nu)|_h \le  \sqrt{\frac{-2(n-1)\Lambda}{n}}
 \ee
 (see~\cite[Remark~4.8]{ChHerzlich} for a discussion of \eq{traps} when $k(\nu)=0$).

Without loss of generality~\cite{LawsonMichelsohn}, we can assume
that the spinor bundle $\Spb$ is equipped with a Hermitian product
$\langle\cdot,\cdot\rangle$  such that   Clifford multiplication by
vectors tangent to $\hyp$ is an anti-Hermitian endomorphism. In the
construction we will need  a bundle isomorphism $\gaz:\Spb\to\Spb$
with the following properties:
 \minilab{propx1}\begin{equs}\label{prop1xa}  &
(\gaz) ^2 = 1\;,
 \\
 \label{prop1xb}
 &\forall X\in T\hyp \quad  \gaz X \,\cdot= -X \cdot \gaz
  \;,
 \\
 \label{prop1xc}
 &  (\gaz)^\dagger = \gaz
  \;,
  \\
 \label{prop1xd}
 & D\gaz = \gaz D
  \;,
 \end{equs}
where $(\gaz)^\dagger$ denotes the conjugate of $\gaz$ with respect
to the Hermitian product $\langle\cdot,\cdot\rangle$ (by a small
abuse of notation, we shall use this symbol for the inner product on
spin-space in any dimension), and $X\cdot$ denotes Clifford
multiplication by $X$. Such a map always exists if $\Spb$ is
obtained by pulling-back to $\hyp$ a space-time spinor bundle,
provided one has an  externally oriented isometric embedding of
$(\hyp,g)$ in a Lorentzian space-time available. Then the Clifford
product $N\,\cdot$, where $N$ is the field of Lorentzian unit
normals to the image of $\hyp$, has the required properties.
Regardless of whether or not such a map exists, one can
\emph{always} replace $\Spb$ by a direct sum of two copies of
$\Spb$; then, for $X\in T\hyp$, we let $X\cdot$ denote the Clifford
action of $X$ and we set
\minilab{prop2zn}
\begin{equs}
\label{prop2zan} & \gaz (\psi_1,\psi_2) := (\psi_2,\psi_1)\;,
 \\
 \label{prop2zbn}
 &
    X \cdot (\psi_1,\psi_2):= (X\cdot \psi_1,-X\cdot \psi_2)\;,
 \\
 \label{prop2zcn}
 &
   D_X (\psi_1,\psi_2):= (D_X\psi_1,D_X \psi_2)
  \;.
 \end{equs}
One checks that \eq{prop2zbn} defines a representation of the
Clifford algebra of $(\hyp ,b )$ on $\Spb\oplus\Spb$, and that
\eq{prop1} holds.

One use of $\gaz$  is  to construct Killing vectors for the metric
$b$ out of imaginary Killing spinors. Indeed, if $\psi$ is such a
spinor, and $e_i$ is a (locally defined) ON basis of $T\hyp$, then
the vector field
$$
Y= \langle \psi, \gaz e^i \cdot \psi\ \rangle e_i
$$
is a Killing vector of the metric $b$. Furthermore, the pair
$(V,Y)$, where $V=\langle\psi,\psi\rangle$, defines a $b$-KID,  by which we mean a KID of
$(\hyp,b,0)$.

 We can now give our first result:

\begin{Theorem}[Positive charges theorem]
\label{T1} Consider an initial data set $(\hyp,g,K)$ satisfying
the positivity and fall-off conditions \eq{posencond} and
\eq{falloff}, with $(\hyp, g)$ complete, and with finite total
matter energy: $\rho \in L^1(\hyp)$. We assume that either $\hyp$
has no boundary, or $\partial \hyp$ is compact and either \eq{wft}
(changing $K$ to $-K$ if necessary) or \eq{traps} holds. Suppose
that the Riemannian background metric $b$ admits imaginary Killing
spinors in the asymptotic region, with respect to a spin structure
which extends to the interior of $\hyp$. Let $\mcK_0$ be the
subset of the set of $b$-KIDs which are of the form $(\langle
\psi, \psi\rangle,\langle \psi, \gamma^0e^i\cdot \psi\rangle e_i)$
for some $b$-imaginary Killing spinor $\psi$. Then for all
$X=(V,Y)\in \mcK_0$ we have
$$H(V,Y)\ge 0\;,$$ with equality if and only if $\psi$ asymptotes to
an
imaginary Killing spinor of $(\hyp,g,K)$ associated with $\nabla$.
\end{Theorem}

\begin{Remark}
It should be emphasised that the imaginary Killing spinors provided
by Theorem~\ref{T1} are only defined along $\hyp$, and not in an
associated space-time if there is one.
\end{Remark}

\begin{Remark}
The bundle of spinors which is used in the proof is arbitrary. We
will freely make use of this fact in our analysis in subsequent
sections.
\end{Remark}

\proof  We  use a Witten-type argument, as follows. Let
$(\Spb,\langle\cdot,\cdot\rangle)$ be any Riemannian bundle of
spinors over $(\hyp,g)$ with Hermitian product
$\langle\cdot,\cdot\rangle$, such that Clifford multiplication
(which we denote by ``$\cdot$") is anti-Hermitian, and with a map
$\gaz$ satisfying \eq{propx1}.

Given an initial data set $(\hyp,g,K)$,  a vector field $X$, and a
spinor field $\xi$ we set
 \beal{Kprddef}
 K(X) &:=& K_{i}{^j}X^i e_j\,\cdot\;,
 \\
\nabla_{X}\xi & := & D_{X}\xi + \frac{1}{2}K(X)\gaz  \xi\;.
 \eeal{nezcondef} Here $e_i$ is a local orthonormal basis of $T\hyp$; it
 is straightforward to check that \eq{Kprddef} does not depend upon
 the choice of this basis.

The argument now has two main steps. First, one shows existence of a
spinor $\chi$ satisfying a modified Dirac equation,
\bel{modDirac}
 e^j\cdot\Big(
\nabla_j+i\sqrt{{-\Lambda\over 2n(n-1)}}e_j\cdot
\Big)\chi=0\;,
 \ee
and which asymptotes to $\psi$, where $\psi$ is an imaginary Killing
spinor of the background metric. This can be done by rather obvious
modifications of the arguments in~\cite{ChHerzlich},
compare~\cite{Maerten}, see
also~\cite{Herzlich:mass,BartnikChrusciel1} for the treatment of the
boundary terms arising from a non-empty $\partial \hyp$. Let us
simply point out that one of the ingredients of the proof is a
weighted Poincar\'e inequality, established e.g.\
in~\cite{BartnikChrusciel1} for the metrics of interest. This proves
positivity of the boundary integral in the Witten identity.  The
next step is to prove that this boundary integral coincides with the
Hamiltonian $H(V,Y)$. This is done by following the calculations in
\cite{AndDahl} and \cite{ChHerzlich}. We note that the relevant part
of those calculations does not use the explicit form of the
imaginary Killing spinors, but only the equation satisfied by them.
\qed

\section{Spherical conformal infinity}

A preferred set of background Killing vector fields is provided by
those which are $b$-normal to the initial data surface.  The
resulting Hamiltonians are usually interpreted as energies.  In
contradistinction with the asymptotically flat case, where only one
normal background Killing vector field exists, if one assumes that
conformal infinity has spherical space-like sections, then there are
several normal background Killing vector fields. This implies that
there is not a \emph{single} energy, but rather an \emph{energy
functional $M$}. This functional $M$ is uniquely characterised by
$n+1$ numbers $\mv{\mu}$, $\mu=0,1,\ldots,n$, which transform as a
Lorentz covector under asymptotic isometries\footnote{These
isometries are, essentially, characterised by conformal isometries
of the conformal boundary at infinity (in the current case the
sphere).} of $g$, see~\cite{ChNagy,Wang}. (The component $\mv{0}$
coincides with the Abbott-Deser mass under appropriate
restrictions~\cite{ChNagy}.) It follows that the Lorentzian length
of $\mv{\mu}$ is a geometric invariant of $(\hyp,g)$.

We start by reviewing the known $3+1$ results. The
asymptotically-adS-positive-energy theorem implies that $\mv{\mu}$
is causal, future pointing~\cite{GHWSpires,GHHP,Maerten}
(compare~\cite{ChHerzlich,Wang,Zhang:hpet}). If it vanishes, then
$(\hyp,g,K)$ are initial data for anti-de Sitter
space-time.\footnote{In fact, the proof of this in~\cite{Maerten}
contains a gap which we fill, see the proof of
Theorem~\ref{CTrig2IKS}, end of Section~\ref{SNenem} below.}

 Quite generally, one can view the
hyperbolic space as a unit spacelike hyperboloid in $\R^{n+1}$, the
latter equipped with the Minkowski metric. If one assumes that
$\mv{\mu}$ is timelike, after applying an asymptotic isometry to
obtain $\mv{\mu}=(m,0,\cdots,0)$, the background Killing vector
fields tangent to $\hyp$ can now be split into rotations and
``boosts". In space-time dimension four it is customary to define
the rest-frame angular momentum as
$$ j_{(i)}:=H(0,\beta_{(i)})\;,$$ where the $\beta_{(i)}$'s are the
generators of rotations of $S^{2}$, when embedded in $\R^3$:
$$
\beta_{(i)} = \epsilon_{ijk}x^j\partial_k\;.
$$
 The numerical values
of the remaining three Hamiltonians, associated with the vector
fields $C_{(i)}$ of \eq{KID2} below,  generating boost
transformations, will be denoted by $c_{(i)}$. In the
asymptotically flat case the $c_{(i)}$'s have the interpretation
of the centre of mass of the system, and can always be set to zero
by a translation of the coordinates. This freedom does not exist
in the asymptotically adS situation. We will retain the name
\emph{centre of mass} for the vector $\vec c=(c_{(i)})$.

 It does not appear to be widely known that, in $3+1$-dimensions, the positive energy
 theorem for asymptotically adS initial data implies an upper bound on
 the center of mass and the angular momentum in terms of $m$.  This
 should be contrasted with the asymptotically Minkowskian positive
 energy theorem, which bounds the space-momentum in terms of the
 energy, but does not impose constraints either on angular momentum
 or on centre of mass.\footnote{Schoen (seminar at the ESI, summer
 2003) has shown that there is no bound on the ratio $|\vec j|/m$ for
 vacuum initial data sets with $\Lambda=0$.}
 Recall that with our choices so
far the energy-momentum vector $\mv{\mu}$ lies along the time axis.
A rotation of the coordinate system aligns the angular momentum
vector $\vec j$ along the first coordinate axis. One can then rotate
$\vec c = (c_{(i)})$ to lie in the $x$--$y$ plane.
  It is shown
in~\cite{Maerten} that the positivity theorem \ref{T1} implies the
following inequality
\bel{Jin}m\ge
\sqrt{-\Lambda/3}\sqrt{(|j_{(1)}|+|\cv2|)^2+\cv1^2}\;,
 \ee
with vanishing $m$ if and only if the initial data set arises from
anti-de Sitter space-time.\footnote{The normalisations of the
Hamiltonians are a matter of conventions, ours are as follows: the
mass $\mv0$ is the numerical value of the Hamiltonian associated
with the background Killing vector $\partial_ t$ when the background
adS metric is written in the form $-(1-\Lambda r^2/n(n-1))dt^2 +
(1-\Lambda r^2/n(n-1))^{-1}dr^2 +r^2 d\Omega^2$, where $d\Omega^2$
is the unit round metric on the $(n-1)$-dimensional sphere. This
normalisation is convenient for comparison with the $\Lambda=0$
limit. Next, the angular momentum is the numerical value of the
Hamiltonian associated with the rotations of $S^{n-1}$ normalised so
that a rotation by $2\pi$ is the identity. Finally, the center of
mass is normalised to make the right-hand-side of our inequalities
look simple.}

The inequality \eq{Jin} can be rewritten in the manifestly
rotation-invariant form \bel{Jin2}m\ge \sqrt{-\Lambda/3}\sqrt{|\vec
c|^2+|\vec j|^2+2|\vec c \times \vec j|}\;,\ee where $\vec c \times
\vec j$ is the vector product, while $|\vec j|=
\sqrt{\jv1^2+\jv2^2+\jv3^2}$, etc. In particular we have the
striking upper bounds \bel{Jin3}m\ge\sqrt{-\Lambda/3}|\vec
j|\;,\quad m\ge\sqrt{-\Lambda/3}|\vec c|\;.\ee Thus, both the length
of the angular momentum vector and that of the centre of mass vector
are bounded by (a multiple of) the invariant norm of the mass
functional $M$.

The  first inequality in \eq{Jin3} is a familiar condition in the
explicit family of Kerr-adS metrics (see, e.g.,~\cite{HT}). Thus,
the restriction on the range of parameters stemming from the
Kerr-adS family is not a result of our incomplete knowledge of the
set of all solutions, but a necessary property of non-singular
asymptotically adS space-times satisfying the dominant energy
condition.

The above leaves several questions unanswered and suggest others: is
there an equivalent of \eq{Jin2} when $\mv{\mu}$ is null? what
happens if the inequalities are equalities? what if $\Mnmo$ is a
two-dimensional torus? what happens in higher dimensions? In this
work we give partial or complete answers to these questions.

First some notation: from now on, in space-time dimension $n$, we
view the hyperbolic space as the open unit ball $B^n(1)\subset \R^n$
equipped with the metric $b=\backn = \omega^{-2}\delta$, where
$\delta$ is the standard flat metric on $\R^n$, and
$$
 \omega= \frac{1-|x|^2}2\;.
$$
In the obvious spin frame associated with this conformal
representation\footnote{More precisely, we take a spin frame which
projects to the frame $\theta^i=\omega^{-1}dx^i$, and a local
basis of the spinor bundle in which the $\gamma^\mu$'s are
constant matrices.}, the imaginary Killing spinors of $\backn$
take the form
\bel{immKs}
 \psi_u = \omega^{-1/2}(1-i x^k \gamma^k)u 
 \ee
(summation over $k$), where $u$ is a spinor with constant entries,
while the anti-Hermitian matrices $\gamma^k$ with constant entries
satisfy the flat space Clifford relations
$$
\gamma^i \gamma^j + \gamma^j \gamma^i = -2\delta^{ij}\;.
$$
 (The $\psi_u$'s exhaust the space of
imaginary Killing spinors because the map which assigns $u$ to
$\psi_u(0)$ is a bijection). As already mentioned, we will also need
a Hermitian matrix $\gamma^0$, with constant entries, satisfying
$$
(\gamma^0)^2=1\;,\qquad \gamma^0 \gamma^j + \gamma^j \gamma^0 =
0\;.
$$
(If such a matrix does not exist we first make a doubling
construction on the $u$'s as in \eq{prop2zn}.) The KID $(\Nu
,Y^i_u)$ associated to $\psi_u$ takes the form
\beal{KID1}
 \Nu &:=& \langle \psi_u, \psi_u \rangle = 2\Big(|u|^2
 \underbrace{\frac{1+|x|^2}{1-|x|^2}}_{=:V_{(0)}} +\langle u, i\gamma^k
 u\rangle
 \underbrace{\frac{(-2)x^k}{1-|x|^2}}_{=:V_{(k)}}\Big)\;,
 \\
 Y^i_u \partial_i &:=& \langle \psi_u, \gamma^0\gamma^i \psi_u
 \rangle e_i
 \nonumber
 \\
  &=&
 2\langle u, \gamma^0\gamma^k u\rangle \underbrace{\Big(\frac{1+|x|^2}2\delta^i_k -
 {x^i x^k}\Big)\partial_i}_{:=C_{(k)}}+  \frac 12 \langle u, i\gamma^0(\gamma^k\gamma^i-\gamma^i\gamma^k) u\rangle
 \underbrace{(x_k\partial_i-x_i\partial_k)}_{:=\Omega_{(k)(i)}}
 \;.
 \nonumber
 \\
 &&
 \eeal{KID2}
The KIDs $(V_{(\mu)},0)$, $\mu=0,\ldots,n$, together with
$(0,C_{(k)})$, $k=1,\ldots n$, and $(0,\Omega_{(i)(j)})$, $1\le i <
j \le n$, span the space of KIDs of $(B(1),b,0)$. The
$\Omega_{(i)(j)}$'s obviously generate rotations, and  therefore it
is natural to use the name \emph{angular momenta} for the
corresponding global charges; these will be denoted by $\Jv{i}{j}$.
As shown in~\cite{ChNagy,Wang}, the collection of functions
$(V_{(0)},V_{(1)},\ldots,V_{(n)})$, transforms as a Lorentz covector
under conformal isometries of the boundary at infinity. This is at
the origin of the name \emph{energy-momentum vector}, denoted by
$\mv{\mu}$, for the associated charges. As already mentioned at the
beginning of this section, the $C_{(k)}$'s generate Lorentz boosts,
when the hyperbolic space is embedded as a hyperboloid into
$(n+1)$--dimensional Minkowski space; the associated charges will be
denoted by $c_{(k)}$, and called \emph{center of mass}.

 It will be convenient  to
reduce $\Jv{k}{j} $ to a canonical form. As a matrix $\Jv{k}{j} $
is anti-symmetric, so that there exists an ON-frame in which
$\Jv{k}{j} $ is block-diagonal, built out of two-by-two blocks of
the form
\bel{omegadefn}
 \left[\begin{array}{cc} 0 & \omvi \\ -\omvi & 0
\end{array}\right]\;,
\ee
with furthermore a last column of zeros in odd space-dimension.

Our next result is the following:

\begin{Theorem}
\label{Thgg} Let $\hyp$ be spin with \emph{dim}\,$\hyp=n\ge 3$ and
suppose that $\Mnmo=S^{n-1}$ (then the spin structures on $\hyp$
and $\hypext$ are necessarily compatible). Under the remaining
hypotheses of Theorem~\ref{T1}, $\mv{\mu}$ is causal
future\footnote{The notion of causality of $\Mv\mu$ is determined
by a Lorentzian metric with signature $(1,n)$ defined by the group
of isometries of hyperbolic space~\cite{ChNagy}, with ``future"
defined as $\mv0> 0$.} directed, or vanishes. Furthermore,
\begin{enumerate}
\item
\label{t1p1}
In every conformal frame\footnote{Recall that the
decomposition of $g$ as a background plus a correction term
involves a choice, and that two such choices can be related to
each other by a conformal transformation of the conformal boundary
at infinity, plus higher order corrections~\cite{ChNagy}. We use
the term ``conformal frame"  to emphasise the fact that such a
choice has been made.} it holds that
\bel{Mangm} \ell \mv{0}\ge |\omv1|+|\omv2|+\ldots +|\omv\ellek|
 \ee
  where
\bel{elldef} \ell:= \sqrt{-\frac {n(n-1)}{2 \Lambda}}
 \;.
 \ee
\item \label{T1p3}
If $\mv{\mu}$ is null, then the space of $\nabla$--imaginary Killing
sections of $\Spb\oplus \Spb$ over $\hyp$ (as defined in
\eq{prop2zn}) is at least $\dim \Spb$--dimensional.

\item When $\mv{\mu}$ is timelike we also have, in a frame where
$\mv i=0$,
\bel{Mcenterineq}\ell\mv{0}\ge \sqrt{
c_{(1)}^2+ \cdots +c_{(n)}^2}
 \;.
 \ee
\item
 \label{T1pvan}
If $\mv 0$ vanishes in some conformal frame, then all global
charges vanish, and the space-time metric along $\hyp$ is Einstein
with vanishing Weyl tensor.
\item \label{T1p5}
In dimension $5+1$, in a specific frame which will be defined in
the proof below, we have the stronger inequality, which is
optimal:
\bel{Optineq56} \ell m \ge \sqrt{c_{(1)}^2
+c_{(3)}^2+\cv5^2+\omv1^2+\omv2^2+2\sqrt{(\omv1c_{(1)})^2 +
(\omv2c_{(3)})^2 + (\omv1\omv2)^2}}
 \;.
 \ee
\item \label{T1p6}
Inequality \eq{Optineq56} remains valid and optimal in dimension $4+1$ after
setting $\cv5=0$.

\item \label{T1p7}
Similarly \eq{Optineq56} remains valid and optimal in
dimension $3+1$ after setting $\cv5=\omv2=0$, and is then
identical to \eq{Jin2}.
\end{enumerate}
\end{Theorem}
\begin{Remark}
\label{R1}
 \Eq{Optineq56} suggests that in all dimensions the following (non-optimal) inequality should hold
 $$
 \ell \mv0 \ge \sqrt{\sum_i \cv i^2 +\Big( \sum _{i<j}|\Jv ij|\Big) ^2}
 \;.
 $$
%
\end{Remark}

\begin{Remark}
A class of $4+1$ dimensional examples with $m_{(0)} \ne 0$
saturating the bound \eq{Mangm} is given by the metrics in
\cite{GutowskiReall} with $F^I_{\mu\nu}=0$, or the metrics in
\cite{CveticGaoSimon}.
\end{Remark}

\begin{Remark}
We will see in Section~\ref{SNenem} below that, in dimension $3+1$,
under natural hypotheses $\mv \mu$ cannot be null.
\end{Remark}

\proof {To avoid annoying multiplicative factors involving the
dimension and the cosmological constant, all calculations that
follow are done assuming $\Lambda=-n(n-1)/2$, so that the background
hyperbolic metric has all sectional curvatures equal to one. This
can be achieved by a scaling of the metric; the general result is
then obtained by rescaling back.}

We have
 \beaa
H(\Nu ,Y^i_u)&=& 2H\Big(|u|^2 (V_{(0)},0) +\langle u, i\gamma^k
 u\rangle (V_{(k)},0)
 \\&&
 \phantom{xxxx}+ \langle u, \gamma^0\gamma^k u\rangle (0,C_{(k)})
  + \frac 14 \langle u, i\gamma^0(\underbrace{\gamma^k\gamma^j-\gamma^j\gamma^k}_{=:2\gamma^{kj}})
 u\rangle (0,\Omega_{(k)(j)})\Big)
 \\
 &=& 2\Big(|u|^2 \underbrace{H(V_{(0)},0)}_{\mv{0}} +\langle u, i\gamma^k
 u\rangle \underbrace{H(V_{(k)},0)}_{\mv{k}}
 \\&&
 \phantom{xxxx}+ \langle u, \gamma^0\gamma^k u\rangle \underbrace{H(0,C_{(k)})}_{c_{(k)}} + \frac 12 \langle u, i\gamma^0\gamma^{kj}
 u\rangle \underbrace{H(0,\Omega_{(k)(j)})}_{\Jv{k}{j} }\Big)
 \\
 &=& 2\langle u, \underbrace{\Big (\mv{0} + i \gamma^k \mv{k} + \gamma^0 \gamma^k
 c_{(k)} + \frac 12 i\gamma^0\gamma^{kj}
 {\Jv{k}{j} }\Big)}_{=:Q}u \rangle \;.
 \eeaa
By the positivity  theorem \ref{T1} the matrix $Q$ must be positive
semi-definite. Let us explore the consequences thereof.

We start by restricting our considerations to spinors $u$ satisfying
\bel{gamres}
 \gamma^0u = \pm u
 \ee
and $|u|^2=1$ (recall that $\gamma^0$ is Hermitian, and its
eigenvalues are plus or minus one since its square is one). As
$\gamma^i$ anti-commutes with $\gamma^0$, it maps
$(\pm1)$--eigenspinors of $\gamma^0$ to $(\mp 1)$--eigenspinors;
thus $\gamma^i u$ and $\gamma^0\gamma^i u$ are each orthogonal to
$u$. We conclude that, on the eigenspaces of $\gamma^0$, the
following holds
$$
\langle u,Qu \rangle = \langle u,{\Big (\mv{0}  + \frac i
2\gamma^0\gamma^{kj}
 {\Jv{k}{j} }\Big)}u \rangle \;.
$$
For $n=3$ (compare \eq{omegadefn}) we have
\bel{ThreeJ}
 \frac 12 i\gamma^0\gamma^{kj}
 {\Jv{k}{j} } = \omv1i\gamma^0\gamma^1\gamma^2
 \;,
 \ee
while in higher dimensions $2\ellek\le n \le 2\ellek+1$ we can write
\bel{nJ}
 \frac 12 i\gamma^0\gamma^{kj}
 {\Jv{k}{j} } = \omv1i\gamma^0\gamma^1\gamma^2+ \omv2 i\gamma^0\gamma^3\gamma^4+ \ldots
 +\omv\ellek
 i\gamma^0\gamma^{2\ellek-1}\gamma^{2\ellek}
 \;.
 \ee
The matrices $ i\gamma^0\gamma^{2k-1}\gamma^{2k}$ are Hermitian,
with square one, therefore their eigenvalues are plus or minus one.
 We will need the following:

\begin{Lemma}
\label{LemGoodB} For every collection
$\{\epsilon_a\}_{a=0,\ldots,\ellek}$, with $\epsilon_a^2=1$, after
performing a doubling of $\Spb$ if necessary as in \eq{prop2zn},
there exists $u$ satisfying $ \gamma^0 u = \epsilon_0 u$ and
$$
\forall a\ge 1 \qquad i\gamma^0\gamma^{2a-1}\gamma^{2a}u
=\epsilon_a u \;.
$$
\end{Lemma}
\begin{Remark}
The result is wrong without the doubling in general, which can be
seen by taking $n=2$, $\gamma^1=i\sigma^1$, $\gamma^2=i\sigma^2$,
and $\gamma^0=\sigma^3$, where the $\sigma^i$'s are the usual
two-by-two Pauli matrices.
\end{Remark}

\proof The matrix $ i\gamma^{2\ellek -1}\gamma^{2\ellek }$ is
Hermitian, with square one, therefore its eigenvalues are plus or
minus one. The matrix $\gamma^{2\ellek }$ defines a bijection
between the $(+1)$--eigenspace and the $(-1)$--eigenspace, so that
each of these spaces is non-empty. Let $X_\ellek $ denote the
$\epsilon_0\epsilon_\ellek $--eigenspace of $ i\gamma^{2\ellek
-1}\gamma^{2\ellek }$. For $0\le \mu \le 2\ellek -2$ the matrices
$\gamma^\mu$ commute with $i\gamma^{2\ellek -1}\gamma^{2\ellek }$,
which implies that $X_\ellek $ is invariant under their action. For
$\ellek \ge 3$ we repeat this construction to obtain a subspace
$X_{\ellek -1}\subset X_\ellek $ on which $ i\gamma^{2\ellek
-3}\gamma^{2\ellek -2}=\epsilon_0\epsilon_{\ellek -1}$. After
$\ellek $ steps we obtain a space $X_{0}\subset
X_1\subset\ldots\subset X_\ellek $ which is invariant under
$\gamma^0$. If there exists a spinor $u$ in $X_0$ such that
$\gamma^0u=\epsilon_0 u$, the result immediately follows. Otherwise
we double $\Spb$ as in \eq{prop2zn}, we take $\hat u$ to be any
non-zero element of $X_0$, and we set $u=(\hat u, \epsilon_0 \hat
u)$.
 \qed

Let $u$ be given by Lemma~\ref{LemGoodB} with
$\epsilon_a=-\mathrm{sgn} \omv a$. We obtain
$$
0\le \langle u, Q u \rangle = \Big(\mv{0} - |\omv1|-\ldots
-|\omv\ellek|\Big)|u|^2\;,
$$
proving point \ref{t1p1}:
$$
\mv{0} \ge |\omv1|+\ldots +|\omv\ellek|
 \;.
$$
In particular $\mv{0}$ is non-negative. Since conformal
transformations of the sphere at infinity induce Lorentz
transformations of  $\mv{\mu}$  we obtain that $\mv{\mu}$  is causal
future directed, or vanishes. Equality implies that the boundary
integral in the Witten identity vanishes, and the volume integral
shows that $u$ is an imaginary Killing spinor (on $\hyp$) for the
modified connection \eq{nezcondef}.

If $\mv{\mu}$ is timelike we clearly also have
\bel{AngMomn}
  m \ge |\omv1|+\ldots +|\omv\ellek|
 \;,
 \ee
where
$$
m:=\sqrt{|\eta^{(\mu)(\nu)}\mv{\mu} \mv{\nu}|}\;,
$$
with $\eta^{(\mu)(\nu)}=\mbox{\rm{diag}}(-1,+1,\ldots,+1)$, and the
$\omvi$'s in \eq{AngMomn} are the angular momenta in a Lorentz frame
in which $\mv{\mu}$ is aligned along the time axis.

Still assuming timelikeness of  $M:=(\mv{\mu})$, and choosing an ON
frame in which $M$ is aligned along $e_{(0)}$, we now drop the
condition \eq{gamres} and assume that $n=3$. We retain \eq{ThreeJ},
and make a rotation in the $\{e_1,e_2\}$ plane so that $c_{(2)}=0$.
Since the Hermitian matrices $\gamma^0\gamma^1$ and
$i\gamma^0\gamma^1\gamma^2$ commute, and square to one, we can
choose $u_1$ such that $|u_1|^2=1$ and, replacing $\gamma^1$ by
$-\gamma^1$ and $\gamma^2$ by $-\gamma^2$ if necessary,
$$
i\gamma^0\gamma^1\gamma^2 u_1 = u_1\;,\quad \gamma^0\gamma^1 u_1=
u_1 \;.
$$
Set
 \bel{goodbas}
u_2:=\gamma^0\gamma^3 u_1\;, \quad u_3 := \gamma^0\gamma^2 u_1\;,
\quad u_4:= \gamma^0\gamma^3 u_3= -\gamma^3\gamma^2 u_1
 \;.
 \ee
%
From the Clifford relations one easily finds that
$$
\hspace{-1.5cm}\left(\begin{array}{c}
   Qu_1 \\
   Qu_2 \\
   Qu_3 \\
   Qu_4
 \end{array}
 \right)
= \left(
          \begin{array}{cccc}
            m+(c_{(1)}+\omv1) & c_{(3)} &0&0 \\
           c_{(3)} & m-(c_{(1)}+\omv1)&0&0  \\
            0&0& m+(-c_{(1)}+\omv1) & c_{(3)} \\
           0&0 &c_{(3)} & m-(-c_{(1)}+\omv1) \\
          \end{array}
        \right)
        \left(\begin{array}{c}
   u_1 \\
   u_2 \\
   u_3 \\
   u_4
 \end{array}
 \right)
$$
One can further check that the $u_i$'s form an ON basis as follows:
$u_1$ is orthogonal to $u_2$ because both are eigenvectors of the
Hermitian matrix $i\gamma^0\gamma^1\gamma^2 $ with different
eigenvalues. (This can be  verified by inspecting the sign in front
of $\omv1$ in the matrix above.) For the same reason $u_1$ is
orthogonal to $u_4$, and $u_2$ is orthogonal to $u_3$. It remains to
justify orthogonality of the elements of the pair $(u_1,u_3)$,
similarly for $(u_2,u_4)$. These follow from the fact that the first
spinor in each of those pairs is an eigenvector of
$\gamma^0\gamma^1$ with an eigenvalue different from the second one
in the pair.\footnote{If one uses a space of spinors which carries
an irreducible representation of the Clifford algebra, than the
above matrix describes $Q$ completely. Otherwise one can, using
descending induction, find an ON basis in which $Q$ is
block-diagonal, with blocks as above.}

Thus, the $u_i$'s form an ON basis of $\mathrm{Vect}\{u_i\}$, so
that the positivity properties of $Q$, when restricted to this
subspace, can be read off by calculating the eigenvalues of the
matrix above. These are easily found to be
$$
m\pm \sqrt{(c_{(1)}\pm\omv1)^2 + c_{(3)}^2}\;.
$$
In particular we have rederived the property that $Q$ is
non-negative if and only if Maerten's inequality \eq{Jin2} holds.
Furthermore, there will be at least two linearly independent
imaginary Killing spinors if and only if the kernel of $Q$ is at
least two-dimensional. Under the current hypotheses, and assuming an
irreducible representation of the Clifford algebra, this will happen
if and only if
\bel{2dker} c_{(1)} \omv1 = 0 \quad \Longleftrightarrow\quad
\Jv{i}{j}c^j=0 \quad \Longleftrightarrow\quad \vec j \times \vec c =
0\;.
 \ee

We now return to general dimension, also dropping the assumption
that $\mv{\mu}$ is timelike. We use spinors obtained by the
``doubling" technique as in  \eq{prop2zn}; it then follows that the
matrix $Q$ has the following block structure:
\bel{blockdg}
Q= \left(%
\begin{array}{cc}
\mv{0}+i\gamma^k \mv{k} & \underbrace{-\gamma^k c_{(k)}+ \frac i 2 \Jv{k}{\ellek}\gamma^k\gamma^\ellek}_{=:B} \\
  \gamma^k c_{(k)}+ \frac i 2 \Jv{k}{\ellek}\gamma^k\gamma^\ellek &\mv{0}-i\gamma^k \mv{k} \\
\end{array}%
\right)
 \ee
(Positivity of $Q$ when restricted to spinors of the form $(u,0)$
gives immediately that $\mv{\mu}$ is causal future pointing, which
we already know.)

Suppose that $\mv{\mu}$ is null, then there exists a $ (\frac 12
\dim \Spb)$--dimensional space of $u\in\Spb$ such that
$(\mv{0}+i\gamma^k \mv{k})u=0$. Likewise there exists a $ (\frac 12
\dim \Spb)$--dimensional space of $v\in\Spb$ such that
$(\mv{0}-i\gamma^k \mv{k})v=0$. Applying $Q$ to a pair $(u,\lambda
v)$, where $\lambda \in \C$, with such an  $u$ and $v$, we obtain
$$
0\le \langle(u,\lambda v),Q(u,\lambda v)\rangle = 2 \Re \Big(\langle
u, \lambda B v\rangle\Big)\;.
$$
Since $\lambda$ is arbitrary we conclude that $\langle(u, v),Q(u,
v)\rangle=0$. Thus, the space of pairs $(u,v)$ which lead to a
zero Hamiltonian charge $H$ equals at least $\frac 12 \dim
\Spb+\frac 12 \dim \Spb= \dim \Spb$. Witten's identity shows that
each such $u$ leads to an imaginary $\nabla$-Killing spinor of
$(\hyp,g,K)$, section of $\Spb\oplus\Spb$. This proves point
\ref{T1p3}.

Suppose, next, that $\mv{\mu}$ is timelike, and let us use a
conformal frame in which $\mv{k}=0$. Using a spinor of the form
$(u,iu)$ one obtains instead
$$
0\le \langle(u,i u),Q(u, iu )\rangle = 2 \Re \Big(\langle u,
(\mv{0}+i B) u\rangle\Big)= 2 \langle u, (\mv{0}-i\gamma^kc_{(k)})
u\rangle
$$
for all $u$, proving \eq{Mcenterineq}.

To prove point~\ref{T1pvan}, suppose that $\mv0$ vanishes, then
$\mv k=0$ by causality of $\mv \mu$, further $\Jv{k}{\ell}$
vanishes by \eq{Mangm}. Applying \eq{blockdg} to spinors of the
form $(u,\pm v)$, positivity of $Q$  implies $\cv k=0$. Thus $Q$
vanishes, which implies that the space of Killing spinors has
maximal dimension. One concludes that the space-time metric is
Einstein, with vanishing Weyl tensor, along $\hyp$ by the
calculations in~\cite[Section~4]{Maerten}, which are done there
for $n=3$, but remain valid for larger values of $n$.

In order to establish our remaining claims, we describe now an
attempt to obtain a simple form of $Q$ in higher dimensions. While
part of the calculation that follows can be done in any dimension,
we have only been able to carry it out completely in dimensions
$4+1$ and $5+1$. We assume that $\Mv{k}$ is timelike, and we use
an ON frame adapted to $\Mv{k}$ in which \eq{nJ} holds. In each
plane $\mathrm{Vect}\{e_{2j-1},e_{2j}\}$ we further make a
rotation so that $\cv{2j}=0$. Let $\ellek$
  be such that $2\ellek\le n \le 2\ellek+1$, for $1\le
j\le \ellek$ set
$$
B_j:=i\gamma^0\gamma^{2j-1}\gamma^{2j}\;, \quad A_j= \gamma^0
\gamma^{2j-1}\;,
$$
then the $A_i$'s and $B_i$'s are Hermitian, with square one, and
satisfy  the commutation relations
 \bel{comRelAB}
B_i B_j= B_jB_i\;,\quad B_iA_j =\left\{
                                  \begin{array}{ll}
                                    -A_jB_i, & \hbox{$i\ne j$;} \\
                                    A_jB_i, & \hbox{$i=j$,}
                                  \end{array}
                                \right.
                                \;,\quad A_iA_j =\left\{
                                  \begin{array}{ll}
                                    -A_jA_i, & \hbox{$i\ne j$;} \\
                                    A_jA_i, & \hbox{$i=j$.}
                                  \end{array}
                                \right.
 \ee
Changing some of the $\gamma^k$'s to $-\gamma^k$'s if necessary, we
can find a spinor $u$ such that
$$
\forall i \quad B_i u = u\;.
$$
 Setting
$$
u_i:= A_iu\;,
$$
one easily obtains the $B_ju_i$'s using \eq{comRelAB}:
$$
B_ju_i =\left\{
                                  \begin{array}{ll}
                                    -u_i, & \hbox{$i\ne j$;} \\
                                    u_i, & \hbox{$i=j$.}
                                  \end{array}
                                \right.
$$
For $n=6$ we can enlarge $\mathrm{Vect}\{u_0:=u,u_1,u_2,u_3\}$ to a
space which is invariant under the action of the $A_i$'s by adding,
to the generating family $\{u_\mu\}$, the spinors $u_4:=A_1A_2u$,
$u_5:=A_1A_3u$, $u_6:=A_2A_3u$, and $u_7:=A_1A_2A_3u$. It is then
easy to work out the matrix of $Q$ in that basis (by considerations
similar to the ones after \eq{goodbas} one checks that the $u_\mu$'s
form an orthonormal family); we only report the result for
$\omv3=0$; for typesetting reasons we write $b_i$ for $\omv{i}$ and
$a_i$ for $\cv{2i-1}$:
\newcommand{\Mzero}{m}%
$$
\hspace{-2.5cm} \left[ \begin {array}{cccccccc} \Mzero
+b_{{1}}+b_{{2}}&a_{{1
}}&a_{{2}}&a_{{3}}&0&0&0&0\\\noalign{\medskip}a_{{1}}&\Mzero
+b_{{1}}-
b_{{2}}&0&0&-a_{{2}}&-a_{{3}}&0&0\\\noalign{\medskip}a_{{2}}&0
&\Mzero -b_{{1}}+b_{{2}}&0&a_{{1}}&0&-a_{{3}}&0
\\\noalign{\medskip}a_{{3}}&0&0&\Mzero -b_{{1}}-b_{{2}}&0&a_{{
1}}&a_{{2}}&0\\\noalign{\medskip}0&-a_{{2}}&a_{{1}}&0&\Mzero
-b_{{1}}-
b_{{2}}&0&0&a_{{3}}\\\noalign{\medskip}0&-a_{{3}}&0&a_{{1}}&0&
\Mzero -b_{{1}}+b_{{2}}&0&-a_{{2}}\\\noalign{\medskip}0&0&-a_{
{3}}&a_{{2}}&0&0&\Mzero +b_{{1}}-b_{{2}}&a_{{1}}
\\\noalign{\medskip}0&0&0&0&a_{{3}}&-a_{{2}}&a_{{1}}&\Mzero +b_{{1}}+b
_{{2}}\end {array} \right]
$$
%
One can use  {\sc Maple} or {\sc Mathematica} to compute the
eigenvalues of $Q$ without assuming $\omv3=0$, but this does not
lead to useful expressions. However, suppose that $n=5$; after
embedding the five-dimensional Clifford algebra into a six
dimensional one, this form of a \emph{general} $Q$  holds in the
basis above. A {\sc Maple} calculation shows then that the
eigenvalues of $Q$ on this subspace all have multiplicity two, and
are equal to
$$
 \Mzero \pm \sqrt{c_{(1)}^2 + \cv3^2
+\cv5^2+\omv1^2+\omv2^2\pm2\sqrt{(\omv1c_{(1)})^2 + (\omv2\cv3)^2
+ (\omv1\omv2)^2}}
 \;.
$$
This gives \eq{Optineq56}.

Specialising further to $\cv5=0$, a similar argument gives the
inequality for $n=4$, proving  point~\ref{T1p6}; a further
specialisation leads to point~\ref{T1p7}.
 \qed

\subsection{Impossibility of null energy-momentum when $n=3$}
\label{SNenem}

Under the hypotheses of Theorem~\ref{T1}, equality in \eq{Jin2}
leads to the existence of imaginary $\nabla$-Killing spinors on
$\hyp$. We have the following result, which does \emph{not} assume
a spherical conformal boundary:

\begin{Theorem}
\label{Trig2IKS} Let $\dim \hyp =3$, and suppose that $(\hyp,g,K)$
admits a  non-trivial imaginary Killing spinor for the connection
\eq{nezcondef}. Then:
\begin{enumerate} \item The Killing development of $(\hyp,g,K)$ admits  an imaginary Killing
spinor.
\item If there are two linearly independent
such spinors on $\hyp$, then the Killing development of $(\hyp,g,K)$
is vacuum and has vanishing Weyl tensor.
\end{enumerate}
\end{Theorem}

\smallskip

\begin{Remark}
In higher dimensions, the minimal number of Killing spinors which
enforces the vanishing of the Weyl tensor does not appear to be
known. For example, consider a five-dimensional Lorentzian
Einstein-Sasaki manifold (all regular types can be constructed as
$S^1$-bundles over K\"ahler-Einstein manifolds with negative scalar
curvature, see~\cite{Kath2,Baum2,Bohle}). A Lorentzian
Einstein-Sasaki space is not conformally flat and has (if it is
simply connected) at least two linearly independent imaginary
Killing spinors.\footnote{We are grateful to Helga Baum for those
remarks.} In those examples  we can choose $\dim \Spb=4$, leading to
dimension four of the space of doubled imaginary Killing spinors in
point \ref{T1p3} of Theorem~\ref{Thgg}. Restricting to an
irreducible sub-representation of the Clifford algebra will
presumably lead to a two-dimensional space, so that our constraints
on a null $\Mv \mu$ do not exclude such non-trivial geometries. In
fact, five-dimensional examples with a two-dimensional space of
imaginary Killing spinors can be found within the family described
in Section~\ref{sSHde}, with a toroidal Scri; but note that these do
not have a null $\mv\mu$.
\end{Remark}

\smallskip

\noindent {\sc Proof of Theorem~\ref{Trig2IKS}:} For point 1, we
need to show that existence of \emph{space} imaginary Killing
spinors for \eq{nezcondef}, that is spinors satisfying the
following:
\bel{neweq23} \widehat{\nabla}_{X}\psi\equiv D_{X}\psi +
\frac{1}{2}K(X)\cdot\gamma^{0}\psi+
i\sqrt{\frac{-\Lambda}{2n(n-1)}}X\cdot\psi =0\; ,\quad X\in T\hyp\;,
 \ee
necessarily implies
that of \emph{space-time} imaginary Killing spinors in the Killing
development of $(\hyp,g,K)$.

We will prove the result using Dirac spinors on $\hyp$, and
especially their decomposition into two component spinors, which
simplifies the calculations. We   use Greek indices for space-time,
preserving Latin indices for some of the lower dimensional
situations which follow; two-component spinor indices will be
capital Latin indices as usual  (see \cite{PenroseRindler84v1} for
further two-spinor conventions; note, however, the \emph{opposite
signature} of the metric here).

 A space-time imaginary
Killing spinor $\psi$ can then be represented by a pair of spinor
fields $(\alpha_A,\beta_{A'})$ satisfying the following coupled
system of equations (compare~\cite[Section~2]{ParkerTaubes82}):
\begin{eqnarray}
\nabla_{AA'}\alpha_B&=&b\epsilon_{AB}\beta_{A'}\;,\nonumber\\
\nabla_{AA'}\beta_{B'}&=&b\epsilon_{A'B'}\alpha_A\;,\label{sys1}
\end{eqnarray}
where $b$ is a constant (not to be confused with the background
metric of Section~\ref{SGc}), which without loss of generality may
be assumed real, and is then related to the cosmological constant by
$\Lambda=-6b^2$.

Saturation of \eq{Jin2} implies that the data $(\hyp,g,K)$ admits a
spinor field $\psi$ satisfying the projection into $\hyp$ of
(\ref{sys1}), say
\bel{p1} \Pi_{\gamma}^{\alpha}S_{\alpha}=0, \ee
where $\Pi_{\gamma}^{\alpha}$ is the tensor projecting tangentially
to $\hyp$ and $S_{\alpha}$ stands for:
\begin{equation}
S_\alpha:= \left(
  \begin{array}{c}
\nabla_{AA'}\alpha_B-b\epsilon_{AB}\beta_{A'} \\
    \nabla_{AA'}\beta_{B'}-b\epsilon_{A'B'}\alpha_A \\
  \end{array}
\right) \;. \label{sys3}
 \ee
%

Given a solution $(\alpha_A,\beta_{A'})$ of (\ref{sys1}), another
solution is provided  by
$(\overline{\beta}_A,\overline{\alpha}_{A'})$. The two solutions are
linearly independent unless $\alpha_A$ and $\overline{\beta}_A$ are
proportional, say $\alpha_A=f\overline{\beta}_A$ for some function
$f$. In this case, it follows from (\ref{sys1}) that $f$ is a
complex constant, of modulus one, and it can then be absorbed into a
redefinition of $\beta_{A'}$. Thus, given a solution of
(\ref{sys1}), we necessarily have at least a two-dimensional space
of solutions unless we have a solution of
\beq \nabla_{AA'}o_{B}=b\epsilon_{AB}\overline{o}_{A'}. \label{sys2}
\eeq
For reasons which will appear, we shall call this the \emph{null case},
while a solution of (\ref{sys1}) not of this form we call the
\emph{non-null case}.

Assuming that the full (as opposed to \eq{p1}) system (\ref{sys1})
holds,  by commuting derivatives one finds
\[
\psi_{ABCD}\alpha^D=0=\overline{\psi}_{A'B'C'D'}\beta^{D'}\]
\[\phi_{ABA'B'}\alpha^B=0=\phi_{ABA'B'}\beta^{B'}\]
where $\psi_{ABCD}$ is the Weyl spinor, the spinor representing the
Weyl tensor, and $\phi_{ABA'B'}$ is the Ricci spinor, representing
the trace-free part of the Ricci tensor. In the non-null case, when
$\alpha_A$ and $\overline{\beta}_A$ are linearly independent, it
follows from this that the Weyl and trace-free Ricci tensors both
vanish and the space-time is locally anti-de Sitter.

For non-trivial examples, therefore, we need to be in
the null case. From (\ref{sys2}) by differentiating again we obtain
\[\psi_{ABCD}o^D=0=\phi_{ABA'B'}o^B,\]
so that
\[\psi_{ABCD}=\Psi o_Ao_Bo_Co_D\;,\quad \phi_{ABA'B'}=\Phi o_Ao_B\overline{o}_{A'}\overline{o}_{B'}\;,\]
for complex functions $\Psi$ and $\Phi$.

Even in the null case, if there are two linearly-independent such
solutions, we shall again have, locally,  anti-de Sitter space
(since $\psi_{ABCD}$ and $\phi_{ABA'B'}$ cannot take this form for
two independent spinors).

We return now to \eq{p1}. Suppose first that we are in the non-null
case. The vector $X^{\alpha}$ constructed according to
\beq
X^{\alpha}=\alpha^A\overline{\alpha}^{A'}+\overline{\beta}^{A}\beta^{A'}\;,
 \label{kv1}
\eeq
 will give~\cite{Maerten} `Killing Initial Data' at $\hyp$. In
the Killing development of $(\hyp,g,K)$, $X^{\alpha}$ will be a
future-pointing, timelike Killing vector. From (\ref{kv1}) we see,
at $\hyp$,
\bel{kv11} X^{\alpha}X_{\alpha}=-2V\overline{V},\eeq
where
\[V=\alpha_A\overline{\beta}^A.\]
By (\ref{p1}) we have
\bel{kv12} \Pi_\mu^\alpha \nabla_{\alpha}V=\Pi_\mu^\alpha
b(\alpha_A\overline{\alpha}_{A'}-\overline{\beta}_{A}\beta_{A'}),\eeq
which is real so that the imaginary part of $V$, say $I$, is
necessarily a constant along $\hyp$.

Recall that the Lie-derivative of a spinor field $\alpha_A$ along a
Killing vector $L^{\alpha}$ is defined as
\be {\mcL}_L\alpha_A:=L^{\mu}\nabla_{\mu}\alpha_A+\Phi_A{}^M\alpha_M
\label{LieS} \ee
where the symmetric spinor $\Phi_{MN}$ is defined by
\bel{dK0}
\nabla_{\mu}L_{\nu}=\Phi_{MN}\epsilon_{M'N'}+\overline{\Phi}_{M'N'}\epsilon_{MN}\;,
 \ee
see e.g.~\cite[p.~40]{TodHuggett}. For $X^{\alpha}$, from
(\ref{kv1}) and (\ref{p1}) we find, at $\hyp$,
\[\Pi^{\mu}_{\alpha}\nabla_{\mu}X_{\beta}=2b\Pi^{\mu}_{\alpha}(\alpha_{(M}\overline{\beta}_{B)}\epsilon_{M'B'}+\overline{\alpha}_{(M'}
\beta_{B')}\epsilon_{MB}),\]
so that, at $\hyp$ in the Killing development,
\[\nabla_{\alpha}X_{\beta}=2b(\alpha_{(A}\overline{\beta}_{B)}\epsilon_{A'B'}
+\overline{\alpha}_{(A'}\beta_{B')}\epsilon_{AB})
+n_{\alpha}v_{\beta},\]
for some vector field $v_{\beta}$ where $n_{\alpha}$ is the (unit,
timelike) normal to $\hyp$. Symmetrising over  the indices
$\alpha$ and $\beta$
the left-hand-side vanishes, thus so does the right-hand-side, which implies $v_\alpha=0$. %
Thus the derivative at $\hyp$ of $X$ is
\bel{dK1}
\nabla_{\alpha}X_{\beta}=2b(\alpha_{(A}\overline{\beta}_{B)}\epsilon_{A'B'}+\overline{\alpha}_{(A'}\beta_{B')}\epsilon_{AB})\;,
\ee
so that
$$
\Phi_{AB} = 2 b\alpha_{(A}\overline{\beta}_{B)}\;. $$
 We impose
\bel{sys4}\mcL _X\alpha_A-2ibI\alpha_A=0=\mcL
_X\beta_{A'}-2ibI\beta_{A'}\eeq
in the Killing development, with $\alpha_A$ and $\beta_{A'}$ known
on $\hyp$ and $I$ the value of the (constant) imaginary part of $V$
at $\hyp$. This determines the spinors throughout the Killing
development. Note also that now $\mcL_X V=0$ in the Killing
development, so that $I$ equals $\Im V$ throughout. Furthermore, it
follows that
\[\mcL_X(\alpha_{(A}\overline{\beta}_{B)})=0,\]
so that the Lie derivative along $X$ of both sides of (\ref{dK1})
vanishes, and therefore this equation holds throughout the Killing
development.

From (\ref{sys4}), (\ref{dK1}) and (\ref{LieS}), we now have
\[X^{\alpha}S_{\alpha}=0\]
with $S_{\alpha}$ as in (\ref{sys3}), and from (\ref{sys4})
\[\mcL_XS_{\alpha}=2ibIS_{\alpha}.\]
Since $X^{\alpha}$ is transversal to $\hyp$, this with (\ref{p1})
gives $S_{\alpha}=0$ at $\hyp$, and therefore throughout the Killing
development.  Now we have a solution of (\ref{sys1}) in the Killing
development, which is therefore locally
 anti-de Sitter.

The null case is very similar: now we have a solution of
\bel{p4} \Pi_{\gamma}^{\alpha}S_{\alpha}=0, \ee
where this time $S_{\alpha}$ stands for
\beq S_{\alpha} :=\nabla_{AA'}o_{B}-b\epsilon_{AB}\overline{o}_{B'}.
\label{sys3n} \eeq
We define the Killing vector by
\bel{kv13} X^{\alpha}=o^A\overline{o}^{A'}. \eeq
This is a future-pointing null vector (which is why we called this
the null case). Since $V$ is now zero,  (\ref{sys4}) becomes
\[\mcL_Xo_A=0\]
and we proceed as before.

For the derivative of $X$ we find in this case
\bel{dK11}
\nabla_{\alpha}X_{\beta}=bo_Ao_B\epsilon_{A'B'}+b\overline{o}_{A'}\overline{o}_{B'}\epsilon_{AB}.
\ee
It follows that
\bel{p2} X^{\alpha}S_{\alpha}=0 \ee
at $\hyp$, so that again $S_{\alpha}$ vanishes at $\hyp$, but now
$\mcL_XS_{\alpha}=0$. We conclude as required that $S_{\alpha}$
vanishes in the Killing development.

 \qed

We shall say that $(\hyp,g,K)$ are \emph{smooth at infinity} if
the corresponding initial data for the conformally rescaled metric
are smooth at the conformal boundary at infinity.
We have the following corollary of Theorem~\ref{Trig2IKS}:

 \begin{Theorem}
\label{CTrig2IKS} Under the hypotheses of Theorem~\ref{T1} let
$\dim \hyp =3$, assume that the conformal boundary at infinity
$\,\,\dot{\!\! \hyp}$ is a finite collection of spheres, with the
metric satisfying the decay conditions \eq{falloff} in each of the
asymptotic regions. Suppose moreover that the conformally
completed manifold $ \hyp\cup \,\,\dot{\!\! \hyp}$ is compact. If
$(\hyp,g,K)$ is smooth at infinity
then:
 \begin{enumerate}

 \item \label{p2as}
 If $(\hyp,g,K)$ admits two
linearly independent imaginary Killing spinors for the connection
\eq{nezcondef} (which will be true if $\mv 0$ vanishes),  then the
initial data set arises from a hypersurface in anti-de Sitter
space-time.

 \item \label{p1as}
$\mv \mu$ cannot be null.

 \item  \label{p3as}  Equality
 in \eq{Jin2} together with $\vec j
\times \vec c =0$ (equivalently, $\omv1c_{(1)}=0$) occurs   if and
only if $(\hyp,g,K)$ can be obtained from a hypersurface in
anti-de Sitter space-time.
 \end{enumerate}
\end{Theorem}

\begin{Remark}
The condition that $(\hyp,g,K)$ is smooth at infinity ensures
equality of the Witten boundary integral with Ashtekar's formula
for mass in terms of the Weyl tensor, and can be weakened by
working out the differentiability threshold needed for this
equality.
\end{Remark}


\noindent {\sc Proof of Theorem~\ref{CTrig2IKS}:}
\emph{\ref{p2as}}. By Theorem~\ref{Trig2IKS}, or by
point~\ref{T1pvan} of Theorem~\ref{Thgg} if $ \mv 0$ vanishes, the
Weyl and Ricci tensors vanish so that the Killing development is
locally anti-de Sitter. To prove that it is globally anti-de
Sitter, it suffices (compare the arguments
in~\cite[Theorem~1.4]{Maerten})
to prove that it is
geodesically complete%
\footnote{The proof of geodesic completeness of the Killing
development of $(\hyp,g,K)$ in~\cite[Theorem~1.4]{Maerten}
invokes~\cite[Lemma~1.1]{manderson:stationary}. However, that last
lemma is incorrect. A counter-example is given by the domain of
outer communications of an extreme Reissner-Nordstr\"om black
hole.}.
This will be a consequence of the following Lemma, provided that
we can show that its hypotheses are satisfied:

\begin{Lemma}
 \label{Lcompl}
In space-time dimension $n+1\ge 2$, consider a stationary
Lorentzian metric
\bel{gfourn}
\fourg = -\exp(\mu)(dt+ \underbrace{\theta_idx^i\!}_{=:\theta}\,
)^2+\threeg
 \;,
\ee
on $\mcM:=\R_t\times \hyp$, where $\threeg$ is a complete
Riemannian metric on $\hyp$,  with Killing vector $X=\partial_t$
satisfying
\bel{utKv}
 \exp(\mu):=-g(X,X)\ge \varepsilon , \qquad |\theta|_h \le
 \varepsilon^{-1}\;,
\ee
for some constant $\varepsilon>0$. Then $(\mcM,\fourg)$ is
geodesically complete.
\end{Lemma}

\begin{Remark} This lemma together with the remaining arguments of the proof below shows
that, in all dimensions $n\ge 3$, the vanishing of $\mv 0$ implies
that the data set arises from the anti-de Sitter space-time
\emph{whenever} the subspace of KIDs generated by those arising
from spinors contains a  KID satisfying \eq{utKv}. We will show
that this is necessarily true when $n=3$; to generalise our result
to all dimensions one  would need to justify \eq{utKv}  for $n\ge
4$.
\end{Remark}

\proof Let $\Gamma(s)=(t(s),\lambda(s))$ be an affinely
parameterised maximally extended geodesic in $(\mcM,\fourg)$, set
$$
\epsilon:= \fourg(\dot \Gamma, \dot \Gamma) \;, \quad p:=
\fourg(\dot \Gamma, X) =-e^\mu(\dot t+{\theta(\dot \lambda)}) \;,
$$
thus $\epsilon$ and $p$ are constant along $\Gamma$. Hence
$$
\threeg(\dot\lambda,\dot\lambda)= \epsilon + e^{-\mu}p^2 \le C
$$
for some constant $C$, and then
$$
|\dot t| = |e^{-\mu}p + \theta(\dot \lambda)| \le C' \;,
$$
for some other constant $C'$. This implies that for any bounded
interval $I\subset \R$ the closure $\overline{\Gamma(I)}$ of the
image $\Gamma(I)\subset\mcM$ is compact, and completeness of
$(\mcM,\fourg)$ readily follows.
 \qed

Returning to the proof of Theorem~\ref{CTrig2IKS}, since the Weyl
tensor vanishes, it follows e.g.
from~\cite{HollandsIshibashiMarolf} that in each of the asymptotic
regions the Witten boundary term is identically zero.%
\footnote{In~\cite{HollandsIshibashiMarolf} the existence of a
space-time with a global $\Scri$ is assumed, but the calculations
there can be repeated in our context.}
Hence the matrix $Q$ vanishes, and there is an imaginary Killing
spinor $\chi_u$ at $\hyp$ for every choice of imaginary
$b$-Killing spinor $\psi_u$ in each asymptotic end, with $\chi_u$
asymptoting to zero in all the remaining ends. However, the number
of imaginary Killing spinors is at most equal to the number of
imaginary $b$-Killing spinors in one end; it follows that $\hyp$
can only have one asymptotic region.

We show that at least one of these imaginary Killing spinors leads
to a Killing vector which is timelike everywhere on $\hyp$, with
controlled $\mu$ and $\theta$.

By (\ref{sys1}), for any space-time Killing spinor, we have
\bel{mubound}
 \nabla_\alpha(\alpha_A\overline{\beta}^A-\overline{\alpha}_{A'}\beta^{A'})=0,
\ee
so that the imaginary part $\Im V$ of
$V=\alpha_A\overline{\beta}^A$ is constant. By choosing the
Killing spinor so that this is not zero at some point $p$ on
$\hyp$ we ensure that it is nonzero everywhere. But the Killing
vector $X$ of \eq{kv1} has norm
\bel{mubound44}
-^4g(X,X)=2V\overline{V}\geq 2(\Im V)^2>0
 \;,
\ee
so that the Killing vector is everywhere timelike.

By Lemma~\ref{LemGoodB} there exist spinors $u_{\pm,0}$ such that,
in any dimension $n\ge 3$, for $a=1,\ldots,\lfloor n/2 \rfloor$,
$$
|u_{\pm,0}|=1\;, \quad  \gamma^0u_{\pm,0}=u_{\pm,0}\;,\quad
 i\gamma^0\gamma^{2a-1}\gamma^{2a}u_{\pm,0} =\pm u_{\pm,0}
\;.
$$
The calculations in \eq{KID1}-\eq{KID2} show that the resulting
$b$-KIDs lead, respectively, to the following $b$-Killing vectors
$$
\partial_t \pm \Big(\sum_{a = 1}^{l} \Omega_{(2a-1)(2a)}\Big)
 \;.
$$
Adding, we conclude that the vector subspace of $b$-KIDs generated
by imaginary $b$-Killing spinors contains the vector $\partial_t$.

Recall, next, that under the current conditions $(\hyp,g,K)$ has
the maximal number of imaginary Killing spinors. This implies that
for  every $p\in \hyp$ the map which to $u$ assigns the value
$\chi_u(p)$, where $\chi_u$ is the solution of the Witten equation
which asymptotes to $\psi_u$, is a linear bijection. Now, the
equation $\Im V(p)=0$ defines an algebraic variety in the space of
spinors at $p$, the complement of which is open and dense.
Therefore there exists an open and dense set of $u$'s such that
the corresponding $\chi_u$'s will have $\Im V(p) \ne 0$, and
consequently will lead to timelike KIDs. (A KID is called
\emph{timelike} if the associated Killing vector is.)

Set
$$
r = \frac 1 {1-|x|}
 \;,
$$
where $x$ is a coordinate as in \eq{immKs}. It follows from
\eq{KID1}-\eq{KID2}, together with the asymptotics of solutions of
the Witten equation,
that for any $\epsilon>0$ we can choose $u_{\pm,\epsilon}$ so that
$u_{\pm,\epsilon}$ approaches $u_{\pm,0}$ as $\epsilon$ tends to
zero, and the corresponding KIDs
$(V_{\pm,\epsilon},Y_{\pm,\epsilon})$ are timelike, with, for
large $r$,
\bel{VYasym}
V_\epsilon:= V_{+,\epsilon}+ V_{-,\epsilon} =
(1+O(\epsilon))r\;,\quad |\underbrace{Y_{+,\epsilon}+
Y_{-,\epsilon}}_{=:Y_{\epsilon}}|_g =
 O(\epsilon)r
 \;.
\ee
 (Note that a sum of timelike future oriented
KIDs is timelike.)

We consider the Killing development defined by $(V_\epsilon,
Y_\epsilon)$, with $\epsilon$ sufficiently small so that
$V_\epsilon \ge 2r/3$ and $|Y_{\epsilon} |_g\le r/3$: thus $ \mcM$
is $\R_t\times \hyp$ with metric
$$
 ^4g = - V_\epsilon^2 dt^2 + g_{ij}(dx^i+Y^i_\epsilon dt)(dx^j+Y^j_\epsilon dt)
 \;,
$$
with Killing vector $X=\partial_t$.
 Letting
\bel{mubound2}
\exp({\mu}):= -^4g(X,X) = V_\epsilon^2 - |Y_\epsilon|^2_g \ge
\frac {r^2}{3}
 \;,
\ee
we rewrite the space-time metric $^4g$ as in \eq{gfourn},
\bel{gfourm}
^4g = -\exp(\mu)(dt+ {\theta_idx^i} )^2+\threeg
 \;,
\ee
so that
\bel{thetaform}
 \theta_i = -e^{-\mu}g_{ij}Y^j_\epsilon
 \;.
\ee

The asymptotics \eq{falloff} of $g$, together with compactness of
$ \hyp\cup \,\,\dot{\!\! \hyp}$ and the Hopf-Rinow theorem imply
completeness of $(\hyp,g)$. The Riemannian metric $\threeg$ is
related to the initial data metric $g$ by the equation
$$
\threeg_{ij}= g_{ij}+ \exp(\mu)\theta_i\theta_j
 \;,
$$
and, since the last term gives a non-negative contribution on any
given vector, completeness of $(\hyp,\threeg)$ follows from that
of $(\hyp,g)$.

The function $\mu$ is uniformly bounded away from zero by
\eq{mubound}-\eq{mubound44} and, for all $\epsilon$ sufficiently
small, it tends to infinity as one recedes to infinity on
$\hypext$  by \eq{VYasym}.

Finally, from \eq{thetaform} and \eq{VYasym},  $e^{\mu/2}
|\theta|_g$ is uniformly bounded in (each of) the asymptotic
regions,
and the norm with respect to $\threeg$ is equivalent to that with
respect to $g$ (with error terms of order $\epsilon^2$). Since
$\hyp$ is a union of a compact set and one asymptotic  end where
$\theta$ has already been shown to be controlled, a uniform bound
on $ |\theta|_\threeg$ follows. Point \emph{\ref{p2as}} follows
now from Lemma \ref{Lcompl}.

\emph{\ref{p1as}}: By point \emph{\ref{T1p3}} of
Theorem~\ref{Thgg}
 there are at least two linearly independent
Killing spinors, and the result follows from point
\emph{\ref{p2as}} of the current theorem.

\emph{\ref{p3as}}: As pointed out in the paragraph preceding
\eq{2dker} the hypotheses of point \emph{\ref{p2as}} are
satisfied.
 \qed

\medskip

\subsection{Non-existence of black hole solutions saturating the equality, $n=3$}
\label{sSBhs}

Whatever the dimension $n\ge 3$, there exist higher-genus Kottler
black hole space-times with zero Hamiltonian mass. One could
naively think of those as saturating our positivity bounds.
However, it should be borne in mind that, for reasons already
explained, those solutions (as well as any solutions with the same
asymptotic behavior) do not possess imaginary Killing spinors, so
our inequality does not apply.

We wish to show, under a natural supplementary assumption,
non-existence of $(3+1)$-dimensional \emph{black hole} space-times
(not necessarily vacuum),
 with spherical or toroidal conformal infinity, \emph{saturating} the
angular momentum bound, except perhaps for Weyl-flat solutions.
To be precise, in addition to the hypotheses of the positivity
theorem \ref{T1}, we will assume that $\hyp$ is the union of an
asymptotically hyperbolic region $\hypext$ and of a compact set,
with non-empty smooth boundary. Moreover, we suppose that the
space-time $(\mcM,\fourrg)$ is
\emph{not} conformally flat.%
\footnote{This hypothesis is used to prove that the horizon is
 degenerate. It can be replaced by the condition that the
 imaginary Killing spinor extends smoothly across the event
 horizon, as then the associated Killing vector will be causal, which again enforces
 degeneracy.}
The hypothesis that the bound is saturated implies existence of a
Killing spinor, and thus also of the associated Killing vector
which we call $X$, which must be null by the analysis of
Section~\ref{SNenem}. The hypothesis of existence of a black hole
will be encoded in the assumption that the boundary of $\hyp$,
when moved by the flow of the Killing vector field $X$,
forms\footnote{Note that the level sets of $u$ \emph{are} null
hypersurfaces generated by $X$, but with non-compact intersection
with $\hyp$.} a \emph{null hypersurface} $\mcH$. So, assume for
contradiction that a solution satisfying the above exists. Then
$X$ is necessarily tangent to the generators of $\mcH$, with zero
surface gravity since $X$ is null everywhere. It follows from
\eq{dK11} that the solution is static in the sense that
$X^\flat\wedge dX^\flat=0$, where $X^\flat=\fourrg(X,\cdot)$. In
\emph{vacuum} this implies~\cite{CRT} that the horizon has higher
genus topology. But this contradicts~\cite[Theorem~4.1]{GSWW},
showing that vacuum solutions of the above kind are not possible.
Finally, the reader will easily check that the hypothesis that the
space-time is vacuum plays no role in this argument, because the
energy-momentum tensor of the space-time metric is necessarily
proportional to $X\otimes X$, and such a tensor does not affect
those equations in~\cite{CRT} which are relevant to the problem at
hand, so that there are no non-vacuum black holes satisfying these
conditions either.

\subsection{Siklos waves}
\label{sSSw}

 Theorem \ref{CTrig2IKS} shows that for nontrivial
examples  saturating Maerten's inequality we need there to be just
a one-dimensional family of solutions of (\ref{sys2}). Metrics
with this property will be briefly described in  this section. We
have seen in Section~\ref{sSBhs} that, subject to some natural
restrictions, such examples do not include black hole solutions.
In fact, we shall see in Section~\ref{ssSC} that such examples are
not possible at all in three space dimensions if we further assume
that $\Scri$ has spherical cross-sections and is ``large enough".

The ``Lobatchevski plane waves" of Siklos~\cite{SiklosL}, which we
propose to call Siklos waves, are precisely characterised by the
existence of a nontrivial spinor satisfying (\ref{sys2}). Siklos
shows that it is possible to introduce coordinates so that the
metric may be written as
\beq g=\frac{1}{2b^2x^2}(dx^2+dy^2-2dudv-H(u,x,y)du^2)\;.
 \label{met1}
 \eeq
Here $X=\partial/\partial v$. (The signature of (\ref{met1}) is
reversed as compared to~\cite{SiklosL}.) The Weyl and Ricci
spinors are
\bea \phi_{ABA'B'}&=&\Phi o_Ao_B\overline{o}_{A'}\overline{o}_{B'}
 \;,
\\
\psi_{ABCD}&=&\Psi o_Ao_Bo_Co_D \;, \label{psi1} \label{phi1}
 \eea
where $\Phi$  and $\Psi$ are given in terms of $H$ by\footnote{The
multiplicative factor 1/16 in the equation for $\Phi_{22}$
in~\cite[p.~254]{SiklosL} should be 1/4.}
\bea \Phi&=&-b^4x^4(H_{xx}+H_{yy}-2H_{x}/x)\;.
 \label{phi2}\\
\Psi&=&-b^4x^4(H_{xx}-H_{yy}-2iH_{xy})
 \;,
 \label{psi2}
\eea
 The cosmological constant is $\Lambda=-6b^2$ (this is not the
$\Lambda$ of the Newman-Penrose formalism which would be $24b^2$).

The Killing vector $X$ Lie drags the Weyl spinor (since it is a
symmetry) and the Lie derivative defined by \eq{LieS} commutes
with contractions and tensor products, so that from (\ref{psi1})
\beq X^{\alpha}\partial_{\alpha}\Psi=0. \label{kv2} \eeq

If $H$ is zero, then (\ref{met1}) is the metric of anti-de Sitter
space with $\scri$ at $x=0$. A variety of other choices for the
function $H$
  also lead to anti-de Sitter space, in particular a constant, say
  $H=H_0$, as the coordinate transformation
\[dv\rightarrow dV=dv+\frac{1}{2}H_0du\]
demonstrates.

\subsection{Rigidity in the $n=3$ spherical case}
\label{ssSC}

In Section~\ref{SNenem} we have shown that a null $\mv{\mu}$ cannot
occur.  In this section we wish to show that the remaining
possibilities for equality in \eq{Jin2} only occur in anti-de Sitter
space-time, under the supplementary condition\footnote{This
hypothesis can often be removed by using the Killing development.
This is, unfortunately, not the case for the problem at hand because
of the zeros of the Killing vector at $\scri$.}  that the initial
data set arises from a space-time with a conformal completion at
infinity which is ``sufficiently large in time". By this we mean
that the interval of the $t$--coordinate below has length at least
$\pi$.

 A (four-dimensional) space-time
$(\mcM,\;^4g)$ is said to be asymptotically-anti-de Sitter if it is
smoothly conformal to a manifold $\tM$ with boundary $\ptM\equiv
\scri\approx \R\times S^2$, with the usual condition that the
conformal factor $\Omega$, relating the metrics as
$\,^4g=\Omega^{-2} \;{}^4\tilde g$,  vanishes on $\Scri$ precisely
at order one. It is  further assumed that the restriction of
$\,{}^4\tilde g$ to the conformal boundary at infinity equals
\beq
\zh_{ij}dx^idx^j=d\theta^2+\sin^2\theta d\phi^2-dt^2. \label{met4}
\eeq
 It is then possible to
introduce\footnote{In this section, and only in this section, we use
the convention that $x^0=R$; the reader should not confuse this with
a time-coordinate.} coordinates $(R,x^i)$ for $i=1,2,3$ so that the
space-time metric $^4g=g$ can be written in the form
\beq g=\frac{1}{R^2}(dR^2+h_{ij}(R,x^k)dx^idx^j) \;,
\label{met2}
\eeq
with
\beq
h_{ij}=\zh_{ij}(x^k)+O(R^2) \label{met3}
 \;.
\eeq

The metric (\ref{met1}) with $2b^2=1$ and $H=0$ takes this form,
though the metric of anti-de Sitter space is more commonly written
as
\beq
g=d\psi^2+\sinh^2\psi(d\theta^2+\sin^2\theta d\phi^2)-\cosh^2\psi
dt^2, \label{met5} \eeq
when the substitution $R=\mathrm{e}^{-\psi}$ will cast it in the
form of (\ref{met2}).

Our aim now is to show that the metric of the Siklos wave
(\ref{met1}) cannot be written in the asymptotically-anti-de
Sitter form (\ref{met2}) unless it is exactly anti-de Sitter. Our
technique will be, first to obtain an asymptotic form of the
Killing vector $X$ and then to show that the equation (\ref{kv2})
is incompatible with (\ref{met2}) unless $\Psi=0$.

We suppose then that $X$ is a Killing vector for the metric
(\ref{met2}) and write it in the form
\beq
X=A\frac{\partial}{\partial R}+B^i\frac{\partial}{\partial
x^i}.\label{kv3} \eeq
The Killing equation may be written as
\[X^{\gamma}\partial_{\gamma}g_{\alpha\beta}+g_{\alpha\gamma}\partial_{\beta}X^{\gamma}+g_{\gamma\beta}\partial_{\alpha}X^{\gamma}=0
.\]
Substituting from (\ref{met2}) we obtain for the $(00)$ component
of this
\[R\frac{\partial A}{\partial R}-A=0\]
so that
\beq
A=RV(x^i) \label{A1} \eeq
for some $V(x^i)$, to be found. For the $(0i)$ components we find
\[\frac{\partial A}{\partial x^i}+h_{ij}\frac{\partial B^j}{\partial
  R}=0,\]
so that
\beq
B^j(R,x^k)=B^j_0(x^k)+O(R^2). \label{B1} \eeq
Finally, for the $(ij)$ components we find
\[\lie_Bh_{ij}=2Vh_{ij}-RV\frac{\partial}{\partial R}h_{ij}.\]
The leading term in this equation, with what we have already,
requires
\beq
\lie_{B_0}\zh_{ij}=2V\zh_{ij}\;.
 \label{B2}
 \eeq
Thus $B_0$ is a conformal Killing vector on $\scri$, and our next
task is to find these. We proceed as before, by setting
\beq
B^i_0\frac{\partial}{\partial x^i}=\beta\frac{\partial}{\partial
t}+A^a\frac{\partial}{\partial y^a} \label{B4} \eeq
where $a=2,3$ and $(y^2,y^3)=(\theta,\phi)$. This is to be a
conformal Killing vector of the metric (\ref{met4}) which we write
as
\[\zh_{ij}dx^idx^j=\eta_{ab}dy^ady^b-dt^2.\]
The conformal Killing equation (\ref{B2}) may be written in the
form
\[B^{k}_0\partial_{k}\zh_{ij}+\zh_{ik}\partial_{j}B^{k}_0+\zh_{kj}\partial_{i}B^{k}_0=2V\zh_{ij}
\]
from which, as before, we obtain the system of equations
\bea
\frac{\partial\beta}{\partial t}&=&V\label{kv4}\\
\frac{\partial\beta}{\partial y^a}&=&\eta_{ab}\frac{\partial
  A^b}{\partial t}\label{kv5}\\
\lie_A\eta_{ab}&=&2V\eta_{ab}.\label{kv6} \eea
To solve these, we need to know some facts about conformal Killing
vectors on $\mathbf \mathbf S^2$ (which, by (\ref{kv6}),
$A=A^a\partial_a$ is). The general solution of (\ref{kv6}) is of
the form:
\beq
A^a=Z^a-\eta^{ab}\frac{\partial\alpha}{\partial y^b} \label{ckv1}
\eeq
where $Z^a$ is a Killing vector for $\eta$ and $\alpha$ is a
conformal scalar, by which we mean a solution of the equation:
\beq \mcD_a\mcD_b\alpha=-\alpha\eta_{ab} \label{ckv2} \eeq
where $\mcD_a$ is the Levi-Civita covariant derivative for $\eta$.
Thus, from (\ref{kv6}), $V=\alpha$. Next, it now follows from
(\ref{kv5}) that
\[\frac{\partial}{\partial y^a}(\beta+\frac{\partial\alpha}{\partial
  t})=\eta_{ab}\frac{\partial Z^b}{\partial t}.\]
Taking the divergence of this we find that
\beq
\frac{\partial Z^a}{\partial
  t}=0;\;\;\;\beta=\beta_0-\frac{\partial\alpha}{\partial t}
\label{ckv3} \eeq
 for some $\beta_0$ independent of $y^a$. Finally, integrating (\ref{kv4}) over $\mathbf S^2$,
 and noting that $\alpha$ integrates to zero because of the equation $\Delta \alpha =-2 \alpha$, shows
 that $\beta_0$ is actually constant and $\alpha$ satisfies
\beq
\frac{\partial^2\alpha}{\partial t^2}=-\alpha \label{ckv4} \eeq
which is readily solved.

We may write out solutions explicitly by regarding the
$\mathbf{S}^2$ as the unit sphere in $\mathbb{R}^3$ with Cartesian
coordinates $ {\bf X}=(X^{\mathbf i})$, $\mathbf i=1,2,3$. Then
$\alpha$ is linear in $X^{\mathbf i}$ and, taking account of
(\ref{ckv4}), may be written in the form
\beq
\alpha=-({\bf{a\cdot X}})\cos t-({\bf{b\cdot X}})\sin t
\label{alpha1} \eeq
in terms of a pair of constant vectors $\bf{a}$ and $\bf{b}$. By
(\ref{ckv3}) we obtain
\beq
\beta=\beta_0-({\bf{a\cdot X}})\sin t+({\bf{b\cdot X}})\cos t,
\label{beta1} \eeq
while $Z$ is a Killing vector, so that
\beq Z^a\frac{\partial}{\partial y^a}=M_{\mathbf {ij}}X^{\mathbf
i}\frac{\partial}{\partial
  X^{\mathbf j}}
\label{Z1} \eeq
for a constant, antisymmetric matrix $M_{\mathbf{ij}}$ (where
necessary, indices $\mathbf{i,j}$ can be raised or lowered with
$\delta_{\mathbf{ij}}$).

We have found an asymptotic form for the most general Killing
vector of (\ref{met2}). However, we are interested in null Killing
vectors, so that by (\ref{kv3})
\[
 g(X,X):=\frac{1}{R^2}(A^2+h_{ij}B^iB^j)=0,\]
which implies in particular that
\beq \zh_{ij}B^i_0B^j_0=0, \label{B3} \eeq
so that $B_0$ is also null.

From (\ref{B4}) and (\ref{ckv1}) this is the condition
\[-\beta^2+\eta^{ab}(Z_a-\partial_a\alpha)(Z_b-\partial_b\alpha)=0.\]
Substituting into this from (\ref{alpha1}), (\ref{beta1}) and
(\ref{Z1}), we obtain a series of algebraic equations by equating
to zero coefficients of $1$, $\sin t$, $\cos t$ and $\cos 2t$.
These are
\beq
|{\bf{a}}|^2=|{\bf{b}}|^2;\;\;\;{\bf{a\cdot b}}=0 \label{con1}
\eeq
then
\begin{eqnarray*}
\beta_0a_{\mathbf i}+M_{\mathbf{ij}}b_{\mathbf j}&=&0\\
-\beta_0b_{\mathbf i}+M_{\mathbf{ij}}a_{\mathbf j}&=&0
\end{eqnarray*}
so that
\beq M_{\mathbf{ij}}=\epsilon(a_{\mathbf i}b_{\mathbf j}-a_{\mathbf
j} b_{\mathbf i}) \label{A5} \eeq
for constant $\epsilon$, and finally
\[-\beta_0^2+M_{{\mathbf{ ik}}}M_{{\mathbf jk}}X^{\mathbf i}X^{\mathbf j}+|{\bf{a}}|^2-({\bf{a\cdot
  X}})^2-({\bf{b\cdot X}})^2
   =0\;,
   \]
which implies just
\beq
\beta_0=-\epsilon|{\bf{a}}|^2
\label{A6} \eeq
with $\epsilon^2|{\bf{a}}|^2=1$.

We have found the general form of any null Killing vector in any
asymptotically adS space-time, so that $X$ of Section \ref{sSSw}
must have this form, in any Siklos wave which is asymptotically
adS. There are two families depending on the sign of $\epsilon$
and the six real parameters $({\bf a}, {\bf b})$ subject to
(\ref{con1}). Replacing the Killing spinor by a multiple thereof
if necessary, we can without loss of generality assume
$|{\bf{a}}|=1$. All choices are equivalent up to rotation and the
discrete symmetry $t\rightarrow t+\pi/2$. We make the choices
\[a_1=b_2=\epsilon=1\]
with other terms zero, then with ${\bf X}=(\sin\theta\cos\phi,
\sin\theta\sin\phi,\cos\theta)$ we obtain
\beq
\alpha=-\sin\theta\cos(\phi-t) \label{bf1} \eeq
and
\bea
B_0&=&(-1+\sin\theta\sin(\phi-t))\frac{\partial}{\partial
  t}+\cos\theta\cos(\phi-t)\frac{\partial}{\partial
  \theta}\label{bf2}\\
&&+(1-\sin(\phi-t)\csc\theta)\frac{\partial}{\partial
  \phi}\nonumber.
\eea
Now we have the Killing vector $X$, at least asymptotically,  we
need to solve (\ref{kv2}). For this we need the asymptotic form of
$\Psi$ compatible with (\ref{met2}).  We recall some of the
conventions associated with conformal rescaling in asymptotically
adS space-times. The unphysical metric is
\[\tilde{g}_{\alpha \beta}=R^2g_{\alpha \beta}\]
with $R$ and $g_{\alpha \beta}$ as in (\ref{met2}). From
(\ref{kv3}) and (\ref{A1}), we see that the Killing vector $X$
extends to a smooth vector field on $\scri$, and we have
\bel{Keqx}
 \nabla_\mu X_\nu = \frac 2 {R^{3}} \tilde X_{[\mu
}\partial_{\nu]} R + O(R^{-2})\;, \ee
where $\tilde X_\mu:= \tilde g_{\mu\nu}X^\nu$; here and below
$O(R^k)$ refers to components in the coordinate system $(R,x^i)$.
From \eq{dK1}, 
where
  now $b=1/\sqrt{2}$, and \eq{psi1}, the Weyl tensor $W$ equals
\bel{Weyl}
W=A(dX\otimes dX-(*dX)\otimes (*dX))+B(dX\otimes
(*dX)+(*dX)\otimes dX)
 \;,
 \ee
where $*$ is the space-time Hodge-dual, while
\[\Psi=A-iB\] %
 and $\Psi$ is as in \eq{psi1} and \eq{psi2}. Let $N=R\partial_R$ be
 a unit normal to the level sets of $R$.

 Recall that
the electric and magnetic Weyl tensors at a hypersurface with normal
$N$ are defined as $E_{ij}:=W_{i \gamma
j\delta}N^{\gamma}N^{\delta}$ and $B_{i j }:=*W_{i \gamma
j\delta}N^{\gamma}N^{\delta}$ respectively, where $W$ is the Weyl
tensor and $*W$ is its dual. The rescaled electric Weyl tensor,
finite on $\scri$ (see~\cite[Lemma~3.1]{HollandsIshibashiMarolf} )
is, by equation (2.14) of~\cite{AshtekarDas} and by
\eq{Keqx}-\eq{Weyl}
\begin{eqnarray*}
\tilde{E}_{i j }&=&R^{-1}W_{i \gamma j\delta}N^{\gamma}N^{\delta}\\
&=& R^{-5} \left(A \tilde X_i \tilde X_j + O(\Psi R)\right) \; .
 \eeaa
Similarly the rescaled magnetic part of the Weyl tensor, finite on
$\scri$, is
\begin{eqnarray*}
\tilde{B}_{i j }&=&R^{-1}*\!W_{i \gamma j\delta}N^{\gamma}N^{\delta}\\
&=& R^{-5} \left(B \tilde X_i \tilde X_j + O(\Psi R)\right) \; .
 \eeaa
It follows that $\lim_{R\to 0} R^{-5} \Psi$ exists, and
 \bea
 \lim _{R\to 0} \tilde E_{ij}
& =&M\tilde{X}_{i}\tilde{X}_{j}\;,  \label{elec}
 \eea
where $M=\lim_{R\to 0}R^{-5}\Re\Psi$. Up to a multiplicative
factor, $M$ is the integrand for the asymptotically defined,
Ashtekar-Magnon global
charges~\cite{AshtekarDas,AshtekarMagnonadS}. If $M$ is zero on
$\scri$ then all global charges are zero. $M$ is bounded on
$\scri$ away from the zeroes of $X$, but could be singular where
$X$ is zero.

Now from (\ref{kv2})
\[X^\alpha\partial_\alpha(MR^5)=0\;,
\]
so with $X$ given by (\ref{kv3}), (\ref{bf1}) and (\ref{bf2}), we
need on $\scri$ that
\bel{meq1}
B_0^i\frac{\partial M}{\partial x^i}-5\sin\theta\cos(\phi-t)M=0.
 \ee
On the equator \eq{meq1} can be integrated to give
\beq M(t,\theta=\pi/2,\phi)=(1-\sin(\phi-t))^{-5/2}f(\phi+t) \;,
\label{m2} \eeq
for some function $f$ of $\phi+t$. From \eq{elec} one then has
$$
\tilde E_{tt}= (1-\sin(\phi-t))^{-1/2}f(\phi+t) \;,
 $$
  so $f$ vanishes
if a smooth global $\scri$ exists.

 For $\cos\theta\ne 0$
introduce $F$ by
\beq
M=(\cos\theta)^{-5}F(t,\theta,\phi) \label{f1}
 \eeq
then (\ref{meq1}) becomes
\[
(-1+\sin\theta\sin(\phi-t))\frac{\partial F}{\partial
  t}+\cos\theta\cos(\phi-t)\frac{\partial F}{\partial
  \theta}+(1-\sin(\phi-t)\csc\theta)\frac{\partial F}{\partial
  \phi}=0.
\]
i.e. $F$ is constant on the integral curves of the vector field
$B_0$, which we need to consider. We shall find that $\cos\theta$ is
either zero or asymptotic to zero along every integral curve, and
that the components of $X$ are asymptotic to zero along every
integral curve. From the rate at which these quantities vanish, it
will follow from (\ref{elec}) and (\ref{f1}) that, on the curves
with $\cos\theta\neq 0$, $F$ vanishes on $\scri$, and so does $M$,
while it will follow from (\ref{m2}) that $M$ is zero on the curves
with $\cos\theta=0$. Note that the Ashtekar-Magnon mass equals the
boundary term which arises in Witten's positive energy argument
by~\cite{HollandsIshibashiMarolf}. Hence, under the hypotheses of
Theorem \ref{CTrig2IKS}, we can then conclude that the initial data
set can be embedded into anti-de Sitter space-time.

The integral curves of $B_0$ satisfy the system of equations
\bea
\frac{dt}{d\lambda}&=&(-1+\sin\theta\sin(\phi-t))\label{tdot}\\
\frac{d\theta}{d\lambda}&=&\cos\theta\cos(\phi-t)\label{thetadot}\\
\frac{d\phi}{d\lambda}&=&(1-\sin(\phi-t)\csc\theta)\label{phidot}
\eea
in terms of a real parameter $\lambda$ along the curves.

Equating the right-hand-sides to zero, we see that the fixed
points of $B_0$ lie on the curve $\Gamma$ defined by
$\theta=\pi/2$, $\phi-t= 2k\pi+\pi/2$, for integer $k$, which is a
helix on $\scri$.

We first investigate integral curves with constant $\theta$. Any
curve on which $\theta$ is constant must, by (\ref{thetadot}),
have $\cos\theta$ or $\cos(\phi-t)$ vanishing, but in the second
case, by (\ref{tdot}) and (\ref{phidot}), we arrive again at
$\cos\theta=0$. Thus the only integral curves with $\theta$
constant have $\theta=\pi/2$. Going further with these, we find
that $\phi+t$ must be constant on them. Introduce $\gamma$ by
$\phi-t=2\gamma+\pi/2$ then we find an equation for $\gamma$ which
integrates to give
\[
\cot\gamma=2(\lambda_0-\lambda).
\]
for some constant $\lambda_0$, and so
\[1-\sin(\phi-t)=2\sin^2\gamma=2(1+4(\lambda_0-\lambda)^2)^{-1}.\]

Now suppose we have an integral curve with a point where
$\cos\theta\neq 0$. It is straightforward to check that the
following are constants along the integral curve:
\bea
a&:=&\frac{\sin\theta\sin\phi-\cos t}{\cos\theta}\label{a1}\\
b&:=&\frac{\sin\theta\cos\phi+\sin t}{\cos\theta}\label{b1} \eea
and then that
\begin{eqnarray*}
\frac{d}{d\lambda}\left(\frac{\sin
  t}{\cos\theta}\right)&=&a\\
\frac{d}{d\lambda}\left(\frac{\cos
  t}{\cos\theta}\right)&=&b,
\end{eqnarray*}
so that
\bea
\frac{\sin t}{\cos\theta}&=&a\lambda+c\label{a2}\\
\frac{\cos
  t}{\cos\theta}&=&b\lambda+d,\label{b2}
\eea
for constants $c$ and $d$. Squaring and adding these we find
\beq
\sec^2\theta=(a^2+b^2)\lambda^2+2(ac+bd)\lambda+(c^2+d^2).\label{the1}
\eeq
If $a^2+b^2=0$ then $\theta$ would be constant, but we have just
seen that the only integral curves with constant $\theta$ have
$\cos\theta=0$ at all points. Thus $a^2+b^2\neq 0$, but now as
$\lambda$ goes to plus or minus infinity, $\sec\theta$ is
unbounded, so that $\theta$ must tend to $\pi/2$ on each integral
curve on which $\cos\theta$ is not always zero. More precisely,
\eq{the1} shows that $$\theta -\pi/2 \sim \lambda^{-1}\quad
\Longleftrightarrow \quad \cos \theta  \sim \lambda^{-1}$$ at
infinity, in the sense that there exists a constant $C$ such that
$C^{-1}\lambda^{-1}\le \cos\theta\le C\lambda^{-1}$.

We now look at the components of $B_0$ from (\ref{bf2}) along the
integral curve. From (\ref{a1}), (\ref{b1}), (\ref{a2}) and
(\ref{b2}) we have
\[\tan\theta\cos\phi=-a\lambda+(b-c);\;\;\;\tan\theta\sin\phi=b\lambda+(a+d)\]
so that
\bel{equiveq}
\sin\theta\sin(\phi-t)-1=(ad-bc)\cos^2\theta \sim \lambda^{-2} \;.
 \ee
To justify ``$\sim$", we note that the cofficient of
$\lambda^{-2}$ cannot be zero as
\[ad-bc=\frac{a\cos t-b\sin t}{\cos\theta}\]
and if this were zero then $t$ would be constant along the
integral curve, which is inconsistent with (\ref{tdot}). In the
same manner we find that
\begin{eqnarray*}
\cos\theta\cos(\phi-t) &\sim& \lambda^{-2}\;,
\\
\frac{\sin(\phi-t)}{\sin\theta}-1&\sim&\lambda^{-2}\;,
\end{eqnarray*}
so that, by (\ref{bf2}), all components of $B_0$ vanish at the
rate $O(\lambda^{-2})$ {\emph{and no faster}} along the integral
curve. Putting this with (\ref{the1}) we find that
$(\cos\theta)^{-5}X\otimes X$ is $O(\lambda)$ along the integral
curve and therefore $F$, which is constant along the integral
curve, must vanish. Thus $F$ vanishes on $\scri$, therefore so
does $M$ and all the asymptotically-defined momenta.

\section{Ricci-flat conformal infinity}
\label{STi}

So far we have been mostly assuming that Scri has spherical
cross-sections. In this section we collect some results about
alternatives. In section~\ref{sSTi} we will prove an analogue of the
angular-momentum inequality \eq{Jin2} for toroidal Scris; section
\ref{SKcb} discusses some other possibilities. In the remaining two
sections we review some examples saturating the inequality.

 \subsection{Toroidal infinity}
 \label{sSTi}

We suppose that conformal infinity has toroidal topology
$$
\T^{n-1}:=S^1\times\ldots \times S^1
 $$
with a flat metric $\deltak $. The space-time metric
\bel{Kottler} \backgn = -\frac{r^2}{\ell^2} dt^2 +
\frac{\ell^2}{r^2}dr^2+ r^2 \deltak
 \;,
 \ee
where $\ell$ is related to the cosmological constant $\Lambda$ by
the formula $2\Lambda\ell^2=-n(n-1)$, provides a static vacuum
example satisfying all the conditions of the positivity theorem. The
slices $t=\const.$ have complete induced metric, with one
conformally compactifiable end where $r\to \infty$, as well as a
``cuspidal end" where $r\to 0$. The toroidal
Kottler black holes~\cite{Kottler} 
also belong to this class. Note that the coordinate $r$ in
\eq{Kottler} can be rescaled by a constant factor, a subsequent
redefinition of $\deltak$ and of $t$ preserves then the general form
of the metric. A natural way of getting rid of this freedom is to
assume that the volume of $(\ConfInf,\deltak)$ equals $16\pi$.
Alternatively, one can assume that this volume equals one, and
remove the normalisation constant $1/16\pi$ in front of \eq{mi}.

We consider the following, trivial spin structure over $\ConfInf $:
Let $\Spb''$ be a product Hermitian bundle of spinors over $\ConfInf
$, with a representation of the Clifford algebra of $(\ConfInf
,\deltak )$ via anti-Hermitian matrices. On $\ConfInf $ we use
manifestly flat local coordinates $x^a$, $a=1,\ldots,n-1$, ranging
from $0$ to $2\pi$, and we choose a spin frame so that all
connection coefficients vanish, with the Clifford action of parallel
vectors represented by constant matrices.

The Witten-type proof of the positive energy theorem requires
imaginary Killing spinors in the asymptotic region $\hypext$ ``near
conformal infinity"; in the current case this is the region $r\ge
r_0$ for some large $r_0$, with the initial data metric $g$
approaching the space-part of \eq{Kottler} as in \eq{falloff}, and
with $K_{ij}$ approaching $\zK_{ij}=0$ as required there. To
construct those spinors  we first consider
$\Spb'=\Spb''\oplus\Spb''$, the direct sum of two copies of
$\Spb''$, equipped with the direct-sum sesquilinear product
$\langle\cdot,\cdot\rangle_\oplus$:
\bel{prop2.0}\langle(\psi_1,\psi_2),(\varphi_1,\varphi_2)\rangle_\oplus
:= \langle\psi_1,\varphi_1\rangle+
 \langle\psi_2,\varphi_2\rangle\;.\ee
  For $X\in T\ConfInf $ we let $X\cdot$ denote the
Clifford action of $X$ and, similarly to \eq{prop2zn}, for
$\psi_1,\psi_2\in\Spb''$ we set
\minilab{prop2z}
\begin{equs}
\label{prop2za} & \gaz (\psi_1,\psi_2) := (\psi_2,\psi_1)\;,
 \\
 \label{prop2zb}
 &
    X \cdot (\psi_1,\psi_2):= (X\cdot \psi_1,-X\cdot \psi_2)\;,
 \\
 \label{prop2zc}
 &
   D_X (\psi_1,\psi_2):= (D_X\psi_1,D_X \psi_2)
  \;.
 \end{equs}
One checks that \eq{prop2zb} defines a representation of the
Clifford algebra of $(\ConfInf ,\deltak )$ on $\Spb'$. Further
\minilab{prop1z}
\begin{equs}\label{prop1za}  & (\gaz )^2 = 1\;,
 \\
 \label{prop1zb}
 &\forall X\in T\ConfInf \quad  \gaz X \,\cdot= -X \cdot \gaz
  \;,
 \\
 \label{prop1zc}
 &  (\gaz)^\dagger = \gaz
  \;,
  \\
 \label{prop1zd}
 & D\gaz = \gaz D
  \;,
 \end{equs}

Next, it is convenient to pass to yet another direct sum bundle
$\Spb=\Spb'\oplus\Spb'$,  equipped with the direct-sum Hermitian
product which will be denoted by $\langle\cdot,\cdot\rangle_{\oplus
\oplus}$. We define, for $\psi_1,\psi_2\in\Spb'$, $X\in T\Mnmo$ and
$a\in \C$,
\minilab{prop2}
\newcommand{\gan}{\gamma^{n}}%
\begin{equs}
\label{prop2a} & \gan (\psi_1,\psi_2) := (-\psi_2,\psi_1)\;,
 \\
 \label{prop2b}
 &
    (X \cdot +a \gaz)(\psi_1,\psi_2):= ((X \cdot +a \gaz) \psi_1,-(X \cdot +a \gaz) \psi_2)\;,
 \\
 \label{prop2c}
 &
   D_X (\psi_1,\psi_2):= (D_X\psi_1,D_X \psi_2)
  \;.
 \end{equs}
This provides one more representation\footnote{This representation
will not be irreducible, but this is irrelevant for the positivity
argument. In fact, already the doubling \eq{prop2z} will lead to a
reducible representation of the $(\ConfInf ,\deltak )$--Clifford
algebra extended by adding $\gamma^0$ when $n$ is odd.}
of the Clifford algebra of $(\ConfInf ,\deltak )$, on $\Spb$, with
moreover
\minilab{prop1}
\begin{equs}\label{prop1a}  & (\gan) ^2 = -1\;,
 \\
 \label{prop1b}
 &\forall X\in T\ConfInf\;, \ a\in \C \quad  \gan (X\cdot+a \gaz) = -(X\cdot+ a \gaz)  \gan
  \;,
 \\
 \label{prop1c}
 & (\gan)^\dagger  = -\gan
  \;,
  \\
 \label{prop1d}
 & D\gan = \gan D
  \;.
 \end{equs}

We assume that on $\hypext$ the background metric $b$ takes the
form
\bel{Torbm}
b = (dx^n)^2+ e^{4\mu x^n}\deltak \;;
 \ee
this corresponds to the space-part of the metric~\eq{Kottler} when
$\mu=1/(2\ell)$. The conformal boundary at infinity is constructed
by multiplying by $e^{-4\mu x^n}$, and replacing $x^n$ by
$y=e^{-2\mu x^n}$; the boundary is then the set $\{y=0\}$. We note
that \eq{Torbm} is a complete space-form metric.

 Any  vector $Y\in T\hypext$ can be written in form
$Y=Y^n\partial_n+X^a e_a$, where $e_a=e^{-2\mu x^n}f_a$, and where
the $f_a$'s form a $\deltak$--ON basis. Note that $\{\partial_n,
e_a\}$ form a $b$--ON basis.
%
We define the $b$--Clifford action of $Y$ on $\Spb$ as
$$
Y\cdot = Y^n \gamma^n+X^a f_a\cdot\;.
$$

\newcommand{\ome}[1]{\omega_{#1}}%
 Let the co-frame $\theta^i=(dx^n,\theta^a)$ be dual to
 $(\partial_n,e_a)$, then the only non-vanishing connection
 coefficients are $-\ome{anb}=\ome{nab}=-2\mu \deltak_{ab}$.
One then has
$$
\zD_k = \partial_k - \frac 14 \ome{ijk}e^i\cdot e^j\cdot =
\left\{
  \begin{array}{ll}
    \partial_n , & \hbox{$k=n$;} \\
\partial_b + \mu \gan e_b\cdot , & \hbox{$k=b$.}
  \end{array}
\right.
 $$
It follows  (compare~\cite{Baum}) that for any $\chi \in \Spb'$,
with constant entries, the spinor field
\bel{Spi}
\psi:=\frac {e^{\mu x^n}} {\sqrt{2}}
 (i\chi, \chi)\;,
 \ee
defined over $\hypext$, is an imaginary Killing spinor for $b$; by
definition,
\bel{KiSp} \zD_Y \psi= - \mu  i Y \cdot \psi\;,
 \ee
where $\zD$ denotes the covariant derivative operator of $b$. One
also has
\bel{KiSp4} \forall \ Z \in  T\hypext \qquad \znabla_Z \hat\psi:=
\Big(\zD_Z - \frac 12 \zK_i{^j}Z^ie_j\cdot \gaz\Big)\hat \psi= - \mu
iZ \cdot \hat \psi\;,
 \ee
because the background extrinsic curvature $\zK_{ij}$ of the slices
$t=0$ for the associated space-time background metric $\backgn$
vanishes.

Let $\mcK$ denote the space of  imaginary Killing spinors $ \psi\in
\Gamma \Spb$ constructed so far. As already mentioned in
Section~\ref{SGc}, to any element of $\mcK$ one can associate a KID
of the background initial data $(b,0)$ as follows
\bel{Kvas}
\mcK\ni   \psi \to 
\Big(V=\langle  \psi,  \psi\rangle_{\oplus \oplus}\;, Y=\langle
\psi, \gan\cdot \gaz  \psi\rangle_{\oplus
\oplus}\partial_n+\sum_a\langle \psi, f_a\cdot \gaz
\psi\rangle_{\oplus \oplus}\;e_a\Big)\;.
 \ee
%
Chasing through the definitions we find
\bean
 V &= &\langle  \psi,  \psi\rangle_{\oplus \oplus}= e^{2\mu x^n} \langle \chi,\underbrace{
 \chi}_{=(\chi_1,\chi_2)}
\rangle_{\oplus}
 \\
 &=&
 e^{2\mu x^n}\Big(\langle \chi_1,
\chi_1\rangle+\langle \chi_2, \chi_2\rangle\Big)\;,
 \label{Vdef1}
 \\
 \nonumber
 Y
 &=& \underbrace{\langle
\psi, \gan\cdot \gaz  \psi\rangle_{\oplus \oplus}}_{=0}\partial_n+
 \sum_a\langle  \psi,  f_a \cdot \gaz  \psi\rangle_{\oplus
\oplus}\; e_a
 \\
 \nonumber
 &=&
  e^{2\mu x^n}\sum_a \langle \chi, f_a \cdot \gaz
\chi\rangle_{\oplus}\;e_a =
 \sum_a(\langle \chi_1,
f_a\cdot\chi_2 \rangle - \langle \chi_2, f_a\cdot \chi_1 \rangle)
f_a
 \\
 &=&
 2\sum_a \Re\Big( \langle \chi_1,
f_a\cdot\chi_2 \rangle\Big)f_a \;.
 \eeal{Ydef1}
%

Let $m$ denote the value of $H$ corresponding to the background-KID
$V=e^{2\mu x^n}/\ell$, $Y=0$. This last KID corresponds to the
Killing vector $\partial_t$ of the metric \eq{Kottler}, so that $m$
has the interpretation as energy. Similarly let $\jv b$ be the value
of $H$ corresponding to $f_b$;
thus $V=0$ and $Y^a\partial_a=f_b$. Clearly each $\jv b$ has a
natural interpretation of angular momentum.

Under the hypotheses of Theorem~\ref{T1}, one concludes that the
composition of \eq{Kvas} with the Hamiltonian map \eq{mi} defines a
\emph{positive} Hermitian form on $\mcK$. We have
\beaa H(V,Y) & = & H \Big((\langle \chi_1, \chi_1\rangle+\langle
\chi_2, \chi_2\rangle)(e^{2\mu x^n},0)+2\sum_a\Re( \langle \chi_1,
f_a\cdot\chi_2 \rangle)(0,f_a)\Big)
\\
 & = &
(\langle \chi_1, \chi_1\rangle+\langle \chi_2, \chi_2\rangle)\ell m
+ 2\sum_a\Re( \langle \chi_1, f_a\cdot\chi_2 \rangle)\jv a\ge 0
 \;,
\eeaa
for all constant spinors $(\chi_1,\chi_2)$. This is possible if and
only if\footnote{Indeed, if $|\vec j|=0$ the inequality \eq{Min2} is
clear. Otherwise choose $\chi_2=\sum_a \jv a f_a\cdot \chi_1/|\vec
J|$ to conclude that \eq{Min2} is necessary. The proof of
sufficiency is left to the reader.}
\bel{Min2}
  m\ge \sqrt{-\frac {2 \Lambda}{n(n-1)}} |\vec j|\;, \qquad |\vec j|:=\sqrt{\jv1^2+\ldots+\jv{n}^2}
 \;.
 \ee
We have thus derived the toroidal equivalent of Maerten's inequality
\eq{Jin2}; we emphasise the spin-structure compatibility condition
\eq{H1}.

In space-dimension three \eq{Min2} can be viewed as the special case
$\vec c=0$ of \eq{Jin2}, but the justification of this appears to
require the work above.

Let $\jvn a$ be the angular momentum associated with the Killing
vector $\partial_a$.
It should be clear that with this definition the inequality in
\eq{Min2} remains valid if $|\vec j|$ is taken to be
$\sqrt{\deltak^{ab}\jvn a \jvn b}$, where $\deltak^{ab}$ is the
inverse matrix to $\deltak_{ab}:=\deltak(\partial_a,\partial_b)$.

\subsection{General conformal infinities with parallel spinors}
\label{SKcb}

We now consider a metric \eq{Kottler}, \emph{without} assuming that
$\deltak$ is flat: instead we assume that the manifold
$(\Mnmo,\deltak)$ carries a non-trivial covariantly constant spinor
$\chi$, section of a spinor bundle $\Spb''$. (Such manifolds are
necessarily Ricci flat,
compare~\cite{Bryant2,Ikemakhen,BaumKath,McKWang,McKWang2}.) The
construction of the background imaginary Killing spinors of the
previous section carries over with only trivial modifications to
such a setting. Under the hypotheses of Theorem~\ref{T1} we then
obtain a positive definite quadratic functional on the space of
covariantly constant spinors of $(\Mnmo,\deltak)$. It appears that
an optimal form
of the resulting constraints has to be analysed
case-by-case. Here we only note the following: For every
$\deltak$--parallel $\chi$ the norm squared $\langle \chi,\chi
\rangle$ is constant over $\Mnmo$. It follows that we can normalise
$\chi$ to obtain two KIDs as in \eq{Kvas} with $\chi_2=\pm
\chi_1=\chi$ in \eq{Vdef1}-\eq{Ydef1}, and with  time component of
the associated KIDs equal to one.
 The positivity of
$H$ for both the plus and minus signs then gives
$$
\ell m  \ge |j|\;,
$$
where $j$ is the angular momentum associated with the $b$-Killing
vector $Y$ corresponding to $\chi$, and $\ell$ has been defined in
\eq{elldef}. We thus obtain positivity of $m$, together with an
upper bound on $|j|$ in terms of $m$. The result is optimal if the
space of covariantly constant spinors of $(\Mnmo,\deltak)$ is
one-dimensional. Otherwise we clearly also have the non-optimal
inequality
$$
\ell m  \ge \sup_\psi |j(X_\psi)|\;,
$$
where the supremum is taken over the covariantly Killing spinors
$\psi$ normalised as described above, and $j(X_\psi)$ denotes the
angular momentum along the Killing vector $X_\psi$ associated to
$\psi$.

\subsection{Nonrigidity in the toroidal case for $n=3$}
\label{ssTC}

By Section~\ref{SNenem} equality in \eq{Min2} leads, locally, to
space-forms or to Siklos waves.
 In
order to see that those are compatible with the toroidal topology at
infinity note, first, that the metric (\ref{met1}) with $2b^2=1$ and
$H=0$ gives anti-de Sitter, by introducing $t=(u+v)/\sqrt{2}$ and
$z=(v-u)/\sqrt{2}$:
\[g=\frac{1}{x^2}(dx^2+dy^2+dz^2-dt^2)
 \;.
 \]
This metric covers part of anti-de Sitter space-time. However, we
now impose a periodic identification in $y$ and in $z$. Then this is
a metric with a $\scri$ which is topologically ${\T}^2\times{\R}$ at
$x=0$ and a `hyperbolic cusp' as $x\rightarrow \infty$, as in
\eq{Kottler}. We can retain these asymptotics with a nonzero
$H(u,x,y)$ which is suitably periodic in $u$ and $y$ and decays
appropriately in $x$ as $x$ goes to zero. A simple class of examples may be generated as
follows: take
\[H=xf'(x)-f(x),\]
then from (\ref{phi2}) we find
\beq
\Phi=-\frac{x^6}{4}\left(\frac{f''}{x}\right)'. \label{S1} \eeq
while from (\ref{psi2})
\beq
\Psi=-\frac{x^4}{4}(xf'')' \label{s2} \eeq
For the dominant energy condition  ($T_{ab}v^av^b\ge 0$ for timelike
$v^a$) to hold we need $\Phi$ to be non-negative so set
$\Phi=\frac{x^6}{4}\rho(x)$ for a non-negative function $\rho$. For
simplicity, we assume that $\rho$ has compact support, and then we
solve for $f''$ as
\beq
f''(x)=x\int^\infty_x\rho(y)dy. \label{s3} \eeq
Suppose $\rho$ is supported in $0<a<x<b<\infty$, then so is $\Phi$
(and so also is the energy-momentum tensor). For $x<a$, we find
$f''=mx$ with $m=\int^b_a\rho dx$ and then $\Psi=-mx^5/2$, which is
the rate of decay we found we required in \eq{ssSC}. Thus $\scri$
exists at $x=0$ with this $H$, as with $H=0$.  Letting  $t$ and $z$ be as at the beginning of this section, we require
the level sets of $t$ to be spacelike. This is equivalent to
\bel{space}
 H<2\;,
 \ee
which can  be arranged by the choice of $f(0)$ for any $\rho$ as
above. We note the following formulae for the metric and second
fundamental form of the hypersurface $\hyp:=\{t=0\}$
\beal{KLtwo0}& \displaystyle g_{ij}dx^i dx^j =  {x^{-2}}\Big\{ dx^2
+ dy^2 + \Big(1-\frac H2\Big)dz^2\Big\}\;,\quad \sqrt{\det
g_{ij}}=x^{-3} \sqrt{1-\frac H2} \;, \phantom{xxxx}&
\\
&\displaystyle
  \label{KLtwo01} \quad g_{tt}=-x^{-2}\Big(1+\frac H2\Big)\;, \quad g_{zt}=
x^{-2}\frac H2\;, &
\\
& \label{KLtwo02}\displaystyle K_{ij}dx^i dx^j =-
\frac {H'} {x\sqrt{ 4-2H}}dx\,dz \;, &
\\
&\displaystyle
 |K|_g^2 = \frac{(xH')^2}{2(2-H)^2}\;,
\eeal{KLtwo2}
which shows that $K$ satisfies the decay conditions needed for a
well-defined mass (recall that $\rho$ vanishes near $x=0$).
One can check that
$$
H(x)= H(0)+\int_0^x\frac{y^3}3\rho(y)dy +
\frac{x^3}3\int_x^\infty\rho(y)dy
 \;,
 $$
so that $H$ is a non-decreasing function of $x$ for non-negative
$\rho$ and, subsequently, that \eq{space} will hold if and only if
\bel{space2}
3H(0)+ \int_0^\infty x^3 \rho(x)dx < 6
 \;.
\ee
Assuming this condition, and a compact support of $\rho$ in
$(0,\infty)$, the hypersurface $\hyp=\{t=0\}$ with the induced
fields provides an example of non-trivial initial data set which
saturates the inequality \eq{Min2}, and satisfies all the
hypotheses of the positive energy theorem  in Section~\ref{SGc}.

 For $x>b$, $\Psi=0$ and the space-time is
locally anti-de Sitter. For example, if we choose $H(0)=0=f(0)$
(note that the choice of $f'(0)$ is irrelevant as it does not change
$H$) then for $x>b$, $H$ is constant and equal to
$H_\infty=(\int^b_ax^3\rho dx)/3$. The metric is
\[g=\frac{1}{x^2}(dx^2+dy^2+dz^2-dt^2-\frac{H_\infty}{2}(dt-dz)^2)\]
but the coordinate transformation
\begin{eqnarray*}
d\tilde{z}&=&dz(1-\frac{H_\infty}{4})+\frac{H_\infty}{4}dt\\
d\tilde{t}&=&-\frac{H_\infty}{4}dz+(1+\frac{H_\infty}{4})dt
\end{eqnarray*}
has the effect of setting $H_\infty$ to zero.

Another interesting example is $f= C\sinh x\sin y\sin u$, where $C$
is a constant. Now $H=xf_x-f$ gives vacuum. It satisfies the
asymptotic conditions, with $\Psi=O(x^5)$ for small $x$, but because
the solution is exponentially large for large $x$ the existence of
globally regular spacelike surfaces is not clear.
This leads naturally to the question  of existence of non-trivial
\emph{vacuum}
  initial data sets saturating the angular momentum inequality
  \eq{Min2}.
 Recall that no such black hole solutions exist by the results in
Section~\ref{sSBhs}, but the general result is not known.

\subsection{Higher dimensional examples}
\label{sSHde}
\label{sssHDT} Gibbons and Ruback~\cite{GibbonsRuback}
have presented some metrics which are generalisations of the Siklos
metrics to higher dimensions (compare \cite[p.~14]{Baum2}). In space
dimension $n$ (so space-time dimension $(n+1)$), the metrics can be
written in the form
\beq
g^{GR}=\frac{1}{2b^2x^2}(dx^2+h_{ab}dy^ady^b-2dudv-H(u,x,y^a)du^2)\;.
 \label{met6}
 \eeq
where $h=h_{ab}dy^ady^b$ is a Ricci-flat, Riemannian metric on an
$(n-2)$-dimensional manifold $^{n-2}M$ (compare (\ref{met1})).
From now on we set  $2b^2=1$. To analyse the imaginary Killing
spinor equation we use the frame
$$
\theta^0= \frac {du}x\;, \quad \theta^1=\frac 1 x (dv+\frac H2
du)\;, \quad \theta^2= \frac {dx}x\;, \quad \theta^a=\frac 1 x
\check \theta^a\;,
$$
where $\check\theta^a$ is an ON-frame for $(^{n-2}M,h)$. A somewhat
lengthy calculation shows that
 if $\psi_h$ is a covariantly constant spinor for
$h$ then, in a basis of the spinor bundle where the
$\gamma$--matrices are independent of $x$, $u$ and $v$, the spinor
field
 $\psi=x^{-1/2}\psi_h$ is an
imaginary Killing spinor for (\ref{met6}) and, in fact:

\begin{Proposition}
\label{SpinEx} The  metrics \eq{met6} admit non-trivial imaginary
Killing spinors if and only if $(^{n-2}M, h)$ admits non-zero
covariantly constant spinors.
 \end{Proposition}

So, when such spinors exist,
the volume integral in the Witten identity vanishes, therefore so
does the boundary integral. Assuming the asymptotic conditions
permit the existence of $\scri$,  the metrics \eq{met6} will
therefore saturate the $n$-dimensional version of our bounds.

The Ricci tensor $\Ric^{GR}$ for $g^{GR}$ may be written
\[\Ric^{GR}=-ng^{GR}+2\Phi X\otimes X\]
where $X=\partial/\partial v$ and $\Phi$ is the function
\[\Phi=-
\frac{x^4}4\left(\frac{\partial^2H}{\partial x^2}
-\frac{(n-1)}{x}\frac{\partial H}{\partial
x}+\Delta_hH\right)\;,\]
where $\Delta_h$ is the Laplacian for $h$. The dominant energy
condition again requires $\Phi$ to be positive and as in the three
dimensional case we can readily find solutions with $H$
independent of $u$ and $y^a$: set $\Phi=\frac{1}{8}
x^{n+3}\rho(x)$ then solve
\[(x^{1-n}H'(x))'=-\rho,\]
where prime denotes $d/dx$, to find
\beq\label{gr2} H'(x)=x^{n-1}\int_x^\infty \rho(s)ds, \eeq
(compare (\ref{s3})). Now if we assume that $^{n-2}M$ is compact
and that $u$ is periodic, we obtain a solution with a $\scri$
located at $x=0$, whose cross-sections are $^{n-2}M\times S^1$.
The discussion around (\ref{space}) goes through as before:
$K_{ij}dx^idx^j$ is as in (\ref{KLtwo02}), where now
$x^i=(x,y^a,z)$, and (\ref{KLtwo0}) is replaced by
\[g_{ij}dx^i dx^j =  {x^{-2}}\Big\{ dx^2
+ h_{ab}dy^ady^b + \Big(1-\frac H2\Big)dz^2\Big\}.\]
Therefore, under \eq{space}, and assuming that $\rho$ is
non-negative and compactly supported,  these solutions will satisfy
the global and asymptotic conditions of the positivity theorem.

For the counterpart of (\ref{elec}) we obtain, with conventions as
above and in~\cite{AshtekarDas} and with $R=\frac x {\sqrt 2}$,
\begin{eqnarray*}
\tilde{E}_{ij}&=&\frac{1}{(n-2)}\Big(\frac x {\sqrt 2}\Big)^{2-n}W_{i\alpha j\beta}N^{\alpha}N^{\beta}\\
&=&\frac{-2}{(n-1)}\Big(\frac x {\sqrt
2}\Big)^{2-n}H''\tilde{X}_i\tilde{X}_j
\end{eqnarray*}
which, by (\ref{gr2}), has a finite limit on $\scri$ where $x=0$.

The imaginary Killing spinors  described immediately before the
statement of Proposition \ref{SpinEx} have the property that
\bel{SpinEin}
 X\cdot  \psi =0
\;,
 \ee
where, as before, $X=\langle \psi, \gamma^0\gamma^\mu
\psi\rangle\partial_\mu$, and $\cdot$ denotes the space-time
Clifford multiplication. An analysis similar to that in
Section~\ref{sSBhs} applies, whatever $n\ge 3$, as follows:
Differentiating \eq{SpinEin} one finds, for all $Y$,
$$
(\nabla_Y X)\cdot \psi \sim \psi \;.
$$
By e.g. \cite[Lemma 2.1, point 2]{Leitner} we then have
$$
\nabla_Y X \sim X \;,
$$
which immediately implies staticity:
$$
X_{[\mu}\nabla_{\nu}X_{\rho]}=0\;.
$$
The search for black hole solutions in this class is inconclusive:
Any event horizon would have to be degenerate, and then the
remaining arguments of Section~\ref{sSBhs} show that the
space-metric on the event horizon has a Ricci tensor proportional
to the metric, with negative proportionality constant. The
consequences of this are not clear, as the constraints imposed by
topological censorship~\cite{GSWW} are less stringent in higher
dimensions.
\bigskip

\noindent {\sc Acknowledgements:} We thank the Newton Institute,
Cambridge (PTC, DM, PT),  and the AEI, Golm (PTC, PT), for
hospitality and support during part of work on this paper. We are
grateful to H.~Baum, G.~Gibbons, F.~Leitner, and M.~Stern for
comments and bibliographical advice, and to M-G.~Greuel and S.~Szybka for help with {\sc Singular} and {\sc Maple} calculations.

\def\cprime{$'$} \def\cprime{$'$} \def\cprime{$'$} \def\cprime{$'$}
\providecommand{\bysame}{\leavevmode\hbox
to3em{\hrulefill}\thinspace}
\providecommand{\MR}{\relax\ifhmode\unskip\space\fi MR }
\providecommand{\MRhref}[2]{%
  \href{http://www.ams.org/mathscinet-getitem?mr=#1}{#2}
} \providecommand{\href}[2]{#2}

\bibliographystyle{amsplain}
\bibliography{
../references/newbiblio,%
../references/newbib,%
../references/reffile,%
../references/bibl,%
../references/Energy,%
../references/hip_bib,%
../references/netbiblio,../references/addon}

\end{document}